\documentclass[fleqn,usenatbib]{mnras}

\usepackage{newtxtext,newtxmath}

\usepackage[T1]{fontenc}
\usepackage{ae,aecompl}

\usepackage{graphicx}	
\usepackage{amsmath}	
\usepackage{amssymb}	
\usepackage{threeparttable}

\title[The Peculiar GC Pal 1 in the SDSS-APOGEE Database]{The Peculiar Globular Cluster Palomar 1 and Persistence in the SDSS-APOGEE Database}

\author[F.~Jahandar et al.]
{\parbox{\textwidth}{Farbod Jahandar$^{1}$\thanks{E-mail: \texttt{farbodj@uvic.ca}},
Kim A. Venn$^{1}$,
Matthew D. Shetrone$^{2}$,
Mike Irwin$^{3}$,
Jo Bovy$^{4,5}$\thanks{\texttt{Alfred P. Sloan Fellow}},
Charli M. Sakari$^{6}$,
Collin L. Kielty$^{1}$,
Ruth A. R. Digby$^{1}$ and
Peter M. Frinchaboy$^{7}$}\vspace{0.4cm}\
\\
$^{1}$Department of Physics and Astronomy, University of Victoria, Victoria, BC, V8W 3P2, Canada\\
$^{2}$Mcdonald Observatory, University of Texas at Austin, HC75 Box 1337-MCD, Fort Davis, TX 79734, USA\\
$^{3}$Institute of Astronomy, University of Cambridge, Cambridge, CB3 OHA, UK\\
$^{4}$Department of Astronomy and Astrophysics, University of Toronto, Toronto, ON, M5S 3H4, Canada\\
$^{5}$Center for Computational Astrophysics, Flatiron Institute, 162 5th Ave, New York, NY 10010, USA\\
$^{6}$Department of Astronomy, University of Washington, Seattle, WA, 98195-1580, USA\\
$^{7}$Department of Physics and Astronomy, Texas Christian University, TCU Box 298840, Fort Worth, TX 76129, USA
}

\date{Accepted XXX. Received YYY; in original form ZZZ}

\pubyear{2017}

\begin{document}
\label{firstpage}
\pagerange{\pageref{firstpage}--\pageref{lastpage}}
\maketitle

\begin{abstract}
The SDSS-III APOGEE DR12 is a unique resource to search for stars beyond the tidal radii of star clusters.  We have examined the APOGEE DR12 database for new candidates of the young star cluster Palomar 1, a system with previously reported tidal tails (\citealt{niederste2010tidal}).
The APOGEE ASPCAP database includes spectra and stellar parameters for two known members of Pal 1 (Stars I and II), however these do not agree with the stellar parameters determined from optical spectra by \cite{sakari2011detailed}.   
We find that the APOGEE analysis of these two stars is strongly affected by the known persistence problem (\citealt{Majewski2015}; \citealt{nidever2015data}).
By re-examining the individual visits, and removing the blue (and sometimes green) APOGEE detector spectra affected by persistence, then we find excellent agreement in a re-analysis of the combined spectra. These methods are applied to another five stars in the APOGEE field with similar radial velocities and metallicities as those of Pal 1. Only one of these new candidates, Star F, may be a member located in the tidal tail based on its heliocentric radial velocity, metallicity, and chemistry. The other four candidates are not well aligned with the tidal tails, and comparison to the Besan\c{c}on model 
(\citealt{robin2003synthetic}) suggests that they are more likely to be non-members, i.e.  part of the Galactic halo.  This APOGEE field could be re-examined for other new candidates if the persistence problem can be removed from the APOGEE spectral database.
\end{abstract}

\begin{keywords}
stars: chemical abundances -- techniques: spectroscopy -- globular clusters: individual (Palomar 1) -- globular clusters: tidal tail
\end{keywords}



\section{Introduction}

Palomar 1 (Pal1) is an unusual globular cluster.
It is young (4-6 Gyr; \citealt{sarajedini2007acs}) and it has a high metallicity
([Fe/H] = $-$0.6 $\pm$ 0.1; \citealt{sakari2011detailed}; \citealt{monaco2011high}); however, it 
is located 3.6 kpc above the Galactic plane, and 17.2 kpc  from the
Galactic Centre (\citealt{harris1996catalog}, 2010 edition).
\cite{niederste2010tidal} examined SDSS and HST photometric fields around Pal1, 
and detected a dispersed tidal tail extending up to 1$^o$ ($\sim$ 0.4 kpc, or $\sim$ 80 half-light radii) from either 
side of the  cluster centre, with roughly as many stars in the tails as in the central cluster region. 

Examination of the chemical abundances of the stars in Pal1 can be used to study 
the origin of this system.  
If Pal1 is a globular cluster that has been shredded, then its stars should show a Na-O anti-correlation (\citealt{carretta2010detailed}).   However, if Pal1 is a captured stellar group from a dwarf galaxy, then it can be expected
to show lower ratios of the $\alpha$-elements (amongst other chemical signatures, 
e.g., see \citealt{venn2004stellar}; \citealt{tolstoy2009star}; \citealt{frebel2015near}).  
\cite{sakari2011detailed} determined the elemental abundances of five stars in Pal1 
from  high-resolution HDS Subaru spectroscopy.  There was no evidence for a 
Na-O anti-correlation in the sample, and the [$\alpha$/Fe] ratios were slightly 
lower than Galactic field stars at the same metallicity but only with 1$\sigma$ 
significance.  These signatures do not favour {\it either} scenario for the origin 
of Pal1; however, \cite{sakari2011detailed} also found high values of [Ba/Y] and [Eu/$\alpha$] 
that indicate unique contributions of r-process elements in this system, which seem
to differ from most Galactic stars.   

The physical properties of Pal1 more closely resemble those of young clusters associated with the Sgr stream (i.e. Pal12 and Ter7; \citealt{sakari2011detailed}), or the intermediate-age clusters in the LMC (\citealt{sakari2017chemical}; \citealt{mucciarelli2008chemical}; \citealt{hill2000age}).  Like Pal1, those clusters also have young ages determined from isochrone fitting (\citealt{dotter2008dartmouth}; \citealt{siegel2007acs}; \citealt{salaris2002homogeneous}) and show lower [$\alpha$/Fe] ratios for their metallicities (\citealt{sbordone2007exotic}; \citealt{cohen2004palomar}; \citealt{bonifacio2004sgr}). 
Furthermore, neither Pal1, nor the other young halo clusters, show the
sodium-oxygen anti-correlation that \cite{carretta2010detailed} have shown is
typical of globular clusters in the Milky Way. Another interesting sparse and young cluster in the halo is Rup106. Like Pal1, Rup106 also has low [$\alpha$/Fe] for 
 its metallicity and no Na-O anti-correlation (\citealt{villanova2013ruprecht}). Rup106 is not associated with any stellar streams,
 unlike the Sgr clusters. However, Rup106 also shows low [La/Fe] and [Na/Fe], so does not appear to be directly linked to Pal1.
 Pal1 may also be linked to the Canis Major over-density based on its chemistry, e.g., high [Ba/Fe] and [La/Fe] (\citealt{sakari2011detailed}; \citealt{martin2004canis}; \citealt{chou2010chemical}).

If Pal1 is a tidally disrupted globular cluster, this makes it an excellent 
probe of the shape of the Milky Way halo.  Palomar 5 (Pal5), another low-mass, low-velocity dispersion globular cluster with more spectacular tidal tails, has been used to model the Galactic potential by \cite{bovy2016shape}, \cite{ishigaki2016line} (2016), \cite{grillmair2006detection}, and \cite{belokurov2007orphan}. Pal5 also shows gaps in the tidal tails that have been examined for constraints 
on  mini-halo substructure (\citealt{bovy2016linear}; \citealt{carlberg2012pal}).  
The tidal tails around Pal1 are much shorter. 
Characterizing this system further by identifying member stars in the tidal tails, 
or in a more extended envelope, could be used to better study the shape of the Milky
Way halo and the origin and evolution of this cluster.

In this paper, we examine the SDSS-APOGEE DR 12 database, which targeted Pal1 as part 
of its globular cluster ancillary data project.
Our search for new members of Pal1 
required a critical and substantial re-examination of the individual visit spectra 
and data analysis techniques. In this paper, we present our target selection methods, 
and cleaning of the combined spectra to remove the persistence problem, and re-analysis of the stellar parameters using the FERRE pipeline. We compare the results with those from \cite{sakari2011detailed} and \cite{niederste2010tidal}, as well as with the Besan\c{c}on model (\citealt{robin2003synthetic}).

\begin{figure}
	\includegraphics[width=\columnwidth]{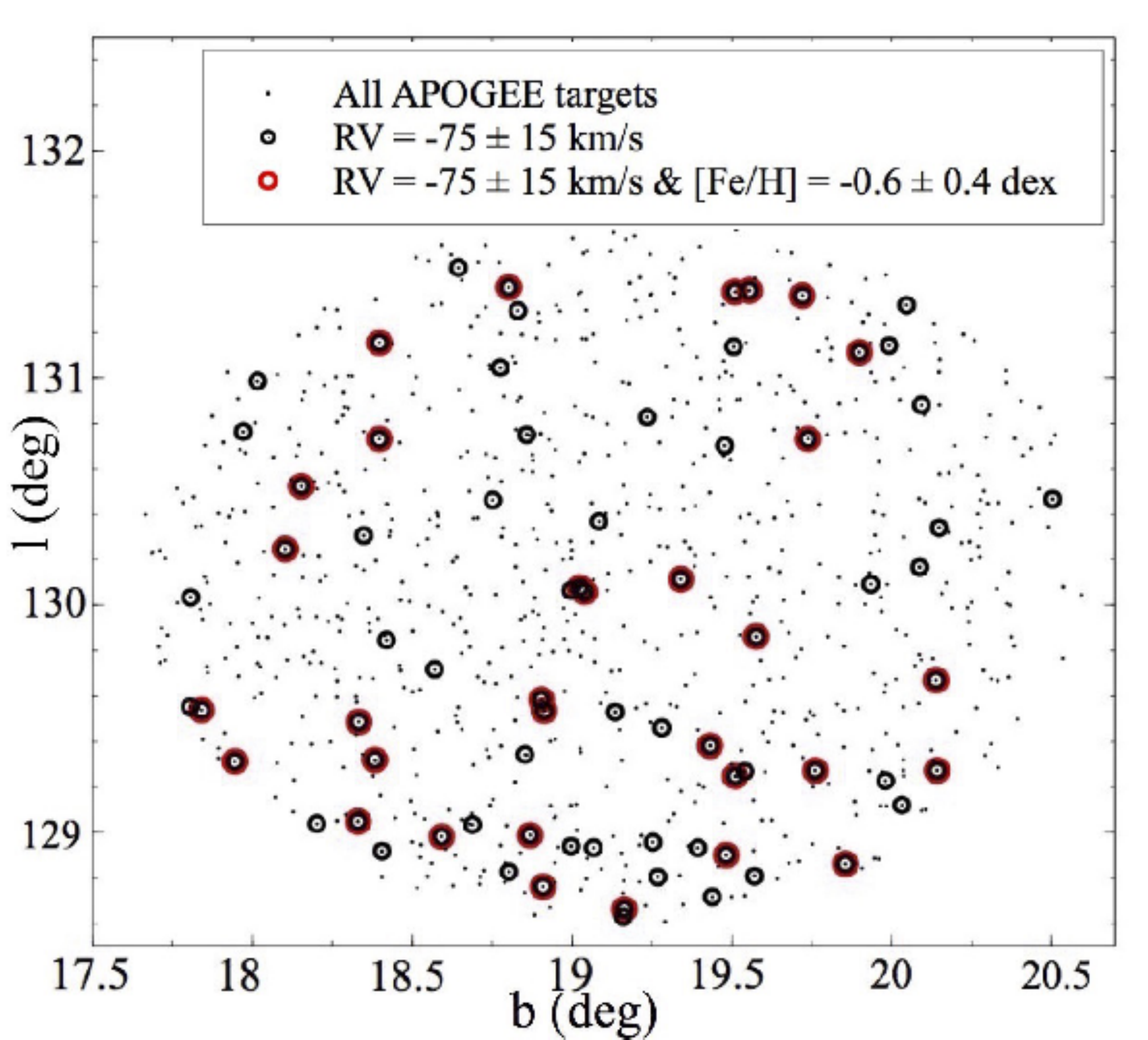}
    \caption{Position of each star from the APOGEE DR12 database in the Pal1 field in Galactic coordinates. Those with heliocentric radial velocities and metallicities similar to those for Pal1 are noted in red circles.}
    \label{fig:Figure1}
\end{figure}



 \begin{figure*}
 \includegraphics[width=\columnwidth]{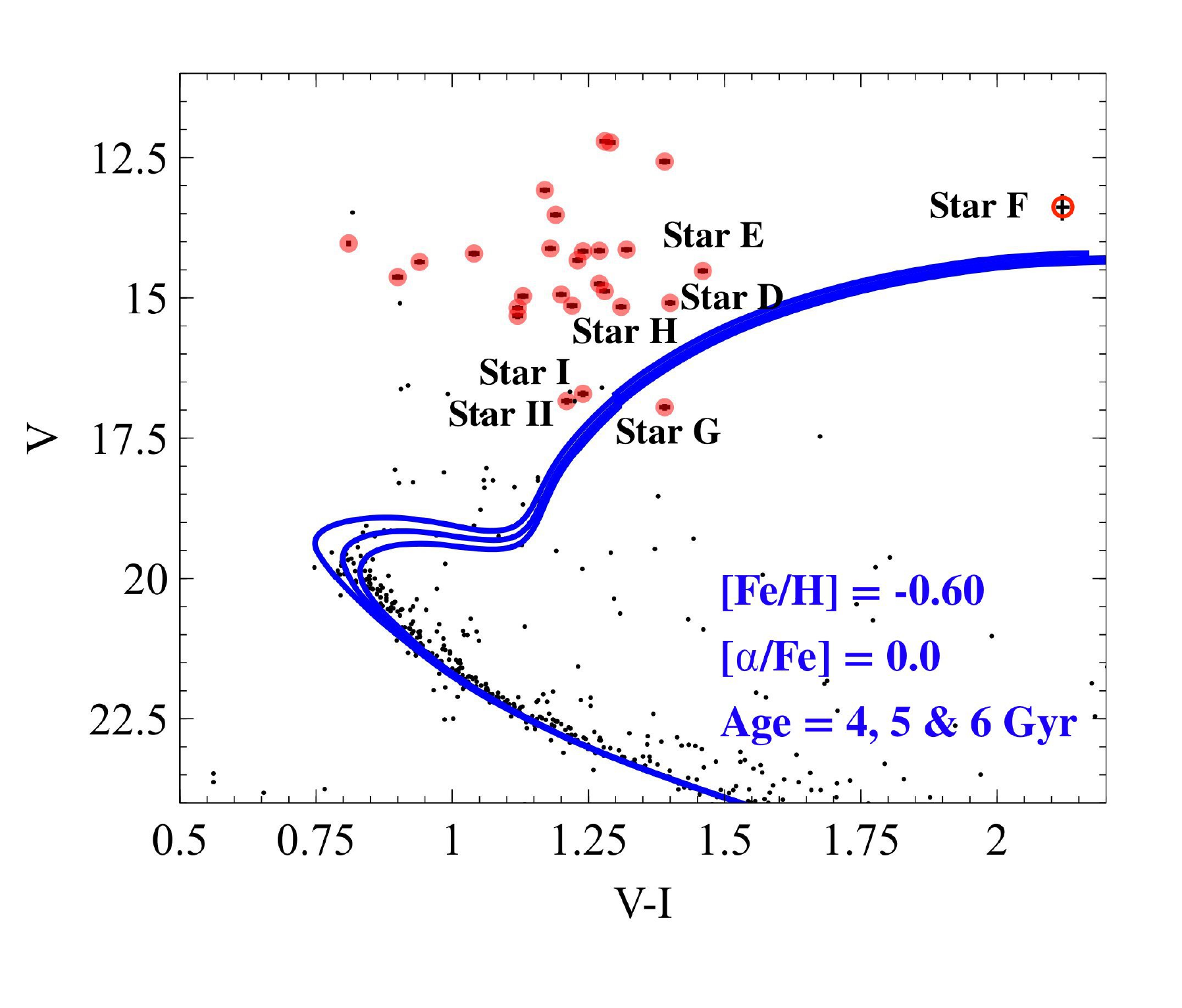}\hfil
    \includegraphics[width=\columnwidth]{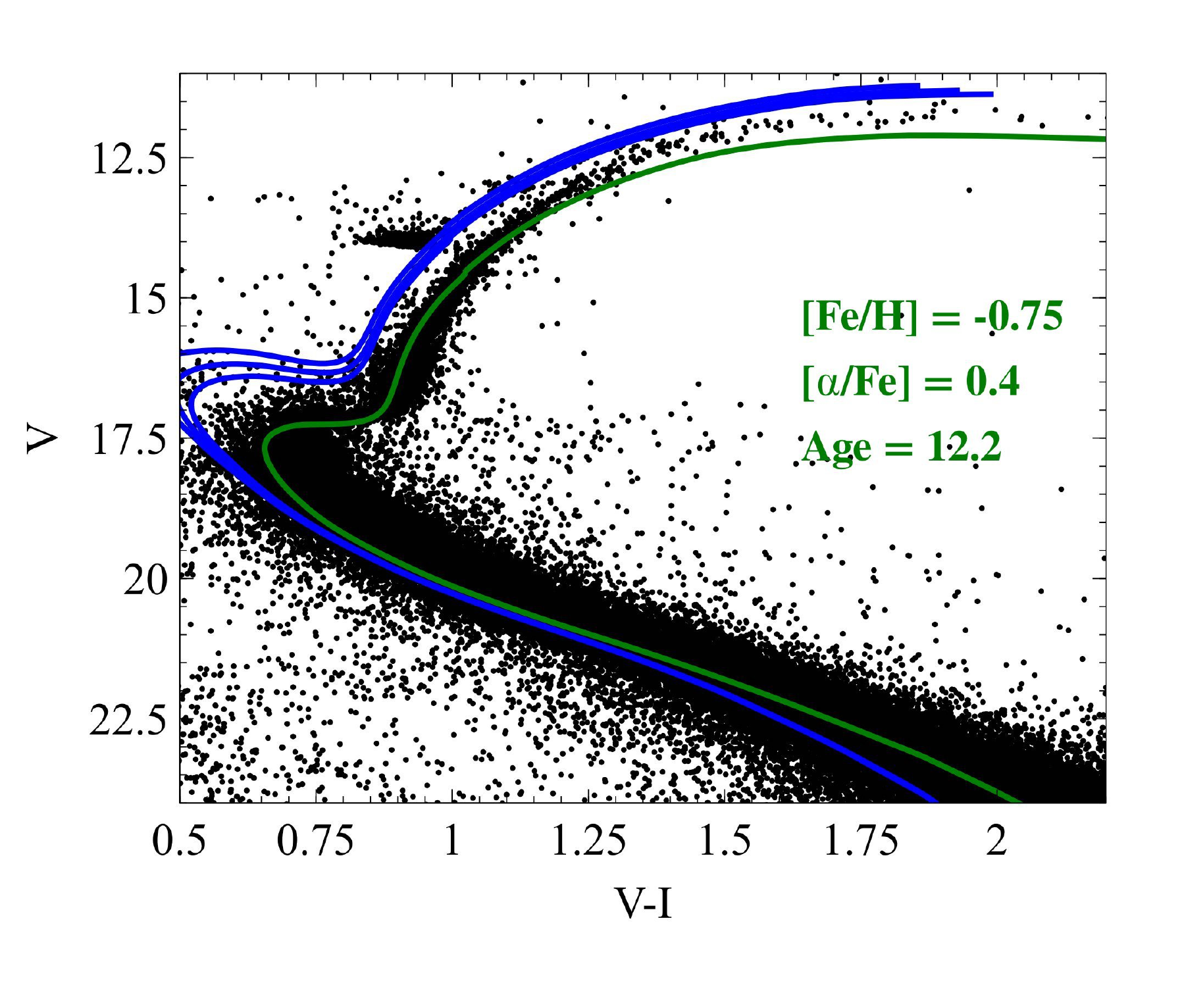}
    \caption{The left panel shows a color-magnitude diagram for Pal~1 (from \citealt{sarajedini2007acs})
with three isochrones for ages 4, 5 and 6 Gyr, from the 
Dartmouth Stellar Evolution Database (\citealt{dotter2008dartmouth}) and the right panel shows the  color-magnitude diagram of 47 Tuc (from \citealt{sarajedini2007acs}) and an isochrone for age 12.2 Gyr as a reference for the position of the red giant branch in a typical globular cluster. Note that the distance modulus and reddening of 47 Tuc is applied to the isochrones of Pal1 in the right panel in order to compare age of the clusters. All of the APOGEE stars with velocities and metallicities similar 
to Pal1 are shown by the red solid circles in the left panel.   The new Pal1 candidate
stars, and Stars I and II, are selected as those closest to the isochrones.
Star F is denoted by an empty red circle because it is flagged by SDSS 
with unreliable photometric magnitudes ("{\it too few detection to be 
deblended}").}
    \label{fig:Figure2}
\end{figure*}

\section{APOGEE data}

The Apache Point Observatory Galactic Evolution Experiment (APOGEE) is a high-resolution, 
high signal-to-noise infrared (IR) spectroscopic survey of over 100,000 red giant stars across 
the full range of the Galactic bulge, bar, disk, and halo (\citealt{Majewski2015}).
The survey was carried out at the 2.5-m Sloan Foundation Telescope 
in New Mexico, covering the wavelength range from 1.5 to 1.7 microns in the H band, 
with spectral resolution R = 22,500 (\citealt{gunn20062}).  The APOGEE Stellar Parameters and Chemical 
Abundances Pipeline (ASPCAP) DR12 (\citealt{perez2016aspcap}) is a data analysis pipeline that produces stellar parameters 
and abundances for 15 different elements 
(C, N, O, Na, Mg, Al, Si, S, K, Ca, Ti, V, Mn, Fe and Ni).

APOGEE uses the same field size and target positioner as the Sloan Extension for 
Galactic Understanding and Exploration (SEGUE) of the Sloan Digial Sky Survey (SDSS).
It uses a series of 7 squared degree tiles to sample the sky with 2" fibres that observe 300 
targets simultaneously.  One of these tiles was centred on Pal1 
(RA =53.33$^o$ \& Dec =79.58$^o$, \citealt{harris1996catalog}, 2010 edition) with fibers allocated to a variety
of targets based on the colours of cool stars (see target selection for the APOGEE program
by \citealt{zasowski2013target}).   Foreground dwarfs are removed 
from our analysis, as well as objects that are unlikely to be associated with Pal1 based on their metallicity and radial velocity.
These include objects with radial velocities outside of $-$75 $\pm$ 15 kms$^{-1}$ and 
metallicities outside of $-1.0 <$[Fe/H] $<-0.2$  (i.e., $4\sigma$ and $2\sigma$ of 
the values for confirmed Pal1 members respectively, e.g., Rosenberg 1998, to account for
errors in the APOGEE metallicities and potential kinematic effects along the tidal tails).
These targets are shown in Fig. \ref{fig:Figure1}, where 9\% of the stars 
in this field may be associated with Pal1. 
Two of these are Stars I and II examined from optical spectra by \cite{sakari2011detailed}.
To further select Pal1 members, we examine a 
colour-magnitude diagram (CMD) of stars in the central portion of Pal1 from HST ACS
photometry (\citealt{sarajedini2007acs}); see Fig. \ref{fig:Figure2}. 
Isochrones are generated from the
Dartmouth Stellar Evolution Database (\citealt{dotter2008dartmouth}) are included with ages 
of 4, 5 and 6 Gyrs, with the distance, reddening, and metallicity from \cite{sarajedini2007acs}, and adopting [$\alpha$/Fe]=0.
However, the APOGEE target selection provides Gunn ugriz and JHK magnitudes of the targets (\citealt{doi2010photometric}),
requiring conversion to Johnson VI.  We have adopted the
calibration from Table 4 of \cite{jordi2006empirical} for Population I stars. \footnote{The uncertainties are determined in quadrature
given the uncertainties for each color index listed in the APOGEE DR12 database and formulae by \cite{jordi2006empirical}.} 

The right panel in Fig. \ref{fig:Figure2} shows the CMD of 47 Tuc  and an isochrone generated from the
Dartmouth Stellar Evolution Database (\citealt{dotter2008dartmouth}) with an age of 12.2 Gyr.  The distance and metallicity are from \citealt{sarajedini2007acs}, with [$\alpha$/Fe]=0.4 and E(B-V)=0 \footnote{The reddening for 47 Tuc of E(B-V)=0.055 from \cite{sarajedini2007acs} does not fit the turn-off well. When no reddening is applied, the fit is better (a lower reddening was similarly found by \citealt{schlafly2011measuring}, E(B-V)=0.03).}. Comparing the CMD of Pal1 to that of 47 Tuc in Fig. \ref{fig:Figure2} clearly shows that Pal1 is younger and more sparsely populated than a typical globular cluster.

The V and I magnitudes from this transformation for Stars I and II are in good
agreement with those from the \cite{sarajedini2007acs}; see Table  \ref{tab:table2}.   
An additional five stars (Stars D, E, F, G and H) with radial velocities and metallicities consistent with Pal1 were selected from near the isochrones.
We examine the stellar properties of these additional five stars below.


\section{Stars I and II}

The stellar parameters for Stars I and II are shown in Table \ref{tab:table2}, from the optical 
analysis by \cite{sakari2011detailed}, and the IR analysis of the APOGEE spectra 
through the ASPCAP pipeline.
These two sets of results are in very poor agreement, with differences of
$\Delta$T$_{\rm eff} \sim 1000$ K and $\Delta$log\,$g \sim$1.0, 
resulting in differences in $\Delta$[Fe/H] $\sim -0.4$.

\begin{table*}
	\caption{DR12 ASPCAP results for members and candidates of Pal1}
	\label{tab:table2}
	\begin{tabular}{ccccccccccccl}
	\hline
     & APOGEE ID         & RA     &     Dec &     RV&     T$_{\rm eff}$  & log$g$     & [Fe/H]         & [$\alpha$/Fe]    & V    &I
     & (S/N)         \\
     &   & (deg) & (deg) & (kms$^{-1}$) & (K) & (dex) &(dex)&(dex)& (mag)  &(mag)
     &  &\\
	\hline

Star D     & 2M03100079+7853325         & 47.503     &     78.892 &     $-$84.2&     4957.7  & 2.69     & $-$0.3            & 0.1    & 15.086 
& 13.861
& 85.3         \\
Star E     & 2M04023010+7935181         & 60.625     &     79.588 &     $-$78.4&     4231.1  & 1.49     & $-$0.7           & 0.1  & 14.522 
& 13.664
& 151.2        \\
Star F      & 2M03354183+7841453        & 53.924     &     78.696 &     $-$84.9&     4847.7  & 2.56     & $-$0.3            & 0.1   & 13.386 
& 11.266
& 377.9        \\
Star G     & 2M03070369+7933134         & 46.765     &     79.554 &     $-$62.0&     4564.0 & 3.05      & $-$0.5            & 0.2    & 16.955 
& 15.562
& 68.7         \\
Star H    & 2M03122767+7927416          & 48.115     &     79.462 &     $-$87.4&     4856.8  & 2.75     & $-$0.3             & 0.1    & 15.163 
& 13.854
& 156.6       \\
Star I     &  2M03332183+7935382          & 53.341    &     79.594 &      $-$75.2 &    5710.9 & 3.47      & $-$0.2            & 0.1  & 16.705 
& 15.461
& 83.6    \\
Star II    &  2M03332960+7934162        & 53.373    &     79.571 &      $-$75.3 &    5602.4 & 3.22      & $-$0.1            & 0.1    & 16.840  
& 15.618
& 67.4      \\

Star I  & (\citealt{sakari2011detailed})& 53.341   & 79.589     &       $-$77.2 &   4800.0 & 2.27      & $-$0.61  & 0.01   & 16.705   & 15.459
& 15   \\
Star II & (\citealt{sakari2011detailed}) & 53.373   & 79.571     &       $-$78.0 &   4750.0 & 2.33      & $-$0.61  & $-$0.10  & 16.675 &15.618
& 15   \\

		\hline
	\end{tabular}
\end{table*}

\begin{figure*}
	\includegraphics[width=2\columnwidth]{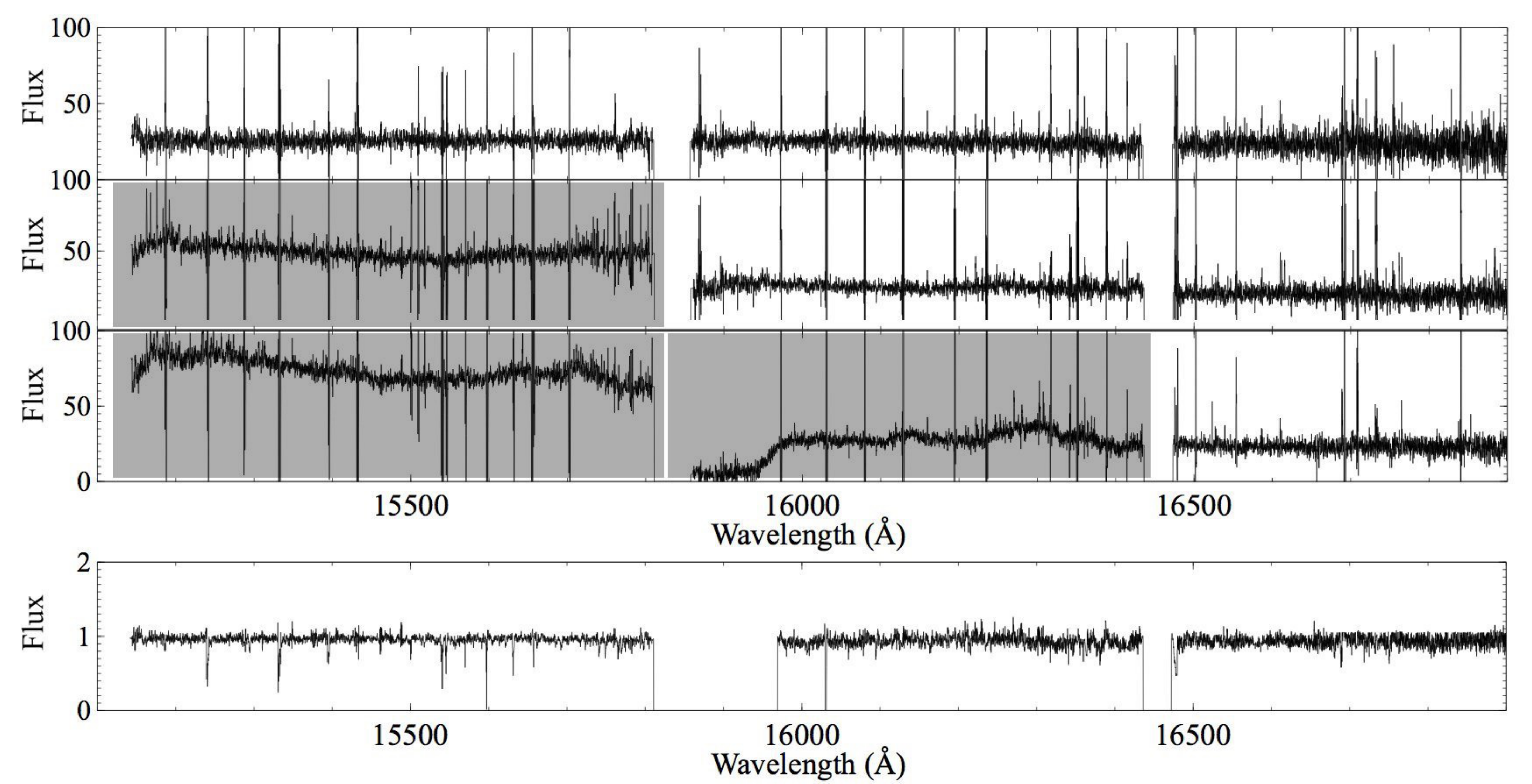}
    \caption{The top three panels are a sample of the spectra with no persistence problems (top), 
moderate persistence (middle), and strong persistence (or other flat fielding problems; bottom).
All chips that would be removed in our analysis are shaded.  
The lower panel shows the final spectra after continuum normalization (see text) and removing sky lines}
    \label{fig:Figure3}
\end{figure*}

\begin{table*}
	\centering
	\caption{Details of each visit for members and candidates of Pal1}
	\label{tab:table3}
		\begin{tabular}{lcccccl} 
		\hline
Visit $\#$ & Visit ID & Heliocentric Velocity 
& Chip A & Chip B& Chip C & S/N\\
        &  & (kms$^{-1})$ & 
        &         &          &        \\
	\hline 
Star I \\
1      & 5283-55816-050 & $-$73.8 & 
P      & \checkmark     & \checkmark     & 10.3 \\
2      & 5282-55822-203 & $-$74.8 & 
\checkmark     & \checkmark     & \checkmark     & 8.5 \\
3      & 5283-55823-050  & $-$71.9& 
P      & P      & \checkmark     & 10.3 \\
4      & 5282-55823-203 & $-$73.7 & 
\checkmark     & \checkmark     & \checkmark     & 15.3 \\
5      & 5282-55841-161 & $-$76.0& 
\checkmark     & \checkmark     & \checkmark     & 13.6 \\
6      & 5283-55843-050 & $-$73.1& 
P      & \checkmark     & \checkmark     & 8.3 \\
7      & 5283-55873-056  &$-$77.2& 
P      & \checkmark     & \checkmark     & 8.2  \\
8      & 5283-55874-056  & $-$75.2& 
P      & \checkmark     & \checkmark     & 15.0 \\
9      & 5283-55905-053  & $-$76.5& 
P      & P      & \checkmark     & 13.6 \\
10     & 5283-55906-053 &$-$75.7
& P      & \checkmark     & \checkmark     & 6.5 \\
11     & 6246-56263-046  & ---
& P      & P      & F   & 9.7 \\
12     & 6247-56264-053  &$-$76.6
& P      & \checkmark     & \checkmark     & 9.8 \\
13     & 6247-56283-044 & $-$76.7
& P      & P      & \checkmark     & 7.3 \\
14     & 6363-56284-183 & $-$75.0&
\checkmark     & \checkmark     & \checkmark     & 9.8 \\
15     & 6364-56285-047  & $-$76.7&
P      & P      & \checkmark     & 10.9 \\
16     & 6246-56539-039  & $-$75.3& 
P      & P      & \checkmark     & 6.1 \\
17     & 6364-56561-038  & $-$72.8& 
P      & P      & \checkmark     & 11.9 \\
18     & 6363-56583-232 & $-$76.4& 
\checkmark     & F   & \checkmark     & 16.8 \\
19     & 6364-56584-032 & ---& 
P      & P      & F & 14.8 \\
20     & 6363-56587-183 & $-$74.0& 
\checkmark     & \checkmark     & \checkmark     & 14.7 \\
21     & 6365-56608-182 & $-$76.0& 
\checkmark     & \checkmark     & \checkmark     & 13.8 \\
22     & 6365-56642-200 &     $-$75.8&
\checkmark     & \checkmark     & \checkmark     & 12.0 \\
23     & 6366-56644-203&     $-$73.8& 
\checkmark     & \checkmark     & \checkmark     & 12.5 \\
24     & 6365-56676-203 &     $-$73.2&
\checkmark     & \checkmark     & \checkmark     & 13.4 \\

Star II \\
1     & 5283-55816-053  & $-$71.9& 
P      & \checkmark  & \checkmark     & 9.1 \\
2     & 5282-55822-204 &$-$75.5& 
\checkmark     & \checkmark     & \checkmark     & 7.1 \\
3     & 5283-55823-053  &$-$71.4& 
P      & P      & \checkmark     & 7.7 \\
4     & 5282-55823-204 & $-$73.4& 
\checkmark     & \checkmark     & \checkmark     & 12.9 \\
5     & 5282-55841-162 & $-$75.1& 
\checkmark     & \checkmark     & \checkmark     & 12.4 \\
6     & 5283-55843-053  &$-$71.1& 
P      & \checkmark     & \checkmark     & 8.4 \\
7     & 5283-55873-059  &$-$73.0& 
P      & \checkmark     & \checkmark     & 7.2 \\
8     & 5283-55874-059  &$-$73.7& 
P      & P      & \checkmark     & 13.8 \\
9     & 5283-55905-050  &$-$74.4& 
P      & P      & \checkmark     & 11.0 \\
10    & 6246-56263-047  &---& 
RV     & RV     & RV     & 7.0 \\
11    & 6247-56264-049  &$-$76.2& 
P      & \checkmark     & \checkmark     & 9.1 \\
12    & 6247-56283-048  &---& 
RV     & RV     & RV     & 9.1 \\
13    & 6363-56284-182 &$-$72.8&
\checkmark     & F   & O      & 8.9 \\
14    & 6364-56285-044  &$-$71.6&
P      & P      & \checkmark     & 7.2 \\
15    & 6364-56561-041  & ---&
RV     & RV     & RV     & 6.8 \\
16    & 6363-56583-233 &$-$74.9& 
\checkmark     & F   & O      & 13.5 \\
17    & 6363-56587-182 &$-$72.1& 
\checkmark     & \checkmark     & \checkmark     & 11.8 \\
18    & 6365-56608-181 &$-$73.9& 
\checkmark     & \checkmark     & \checkmark     & 12.0  \\
19    & 6365-56642-199 &
0& \checkmark     & \checkmark     & \checkmark     & 10.0 \\
20    & 6366-56644-059  &$-$75.0&
P      & P      & \checkmark     & 11.2 \\
21    & 6365-56676-204 &$-$72.2&
\checkmark     & \checkmark     & O      & 11.6  \\

Star D \\
1      & 5282-55815-010  & $-$66.1 &
P       & \checkmark     & \checkmark     & 5.4 \\
2      & 5282-55822-010 & $-$82.6 & 
\checkmark     & \checkmark     & \checkmark     & 21.0 \\
3      & 5282-55823-010  & $-$83.8 & 
\checkmark       & \checkmark       & \checkmark     & 35.8 \\
4      & 5282-55841-004 & $-$87.7 & 
\checkmark     & \checkmark     & \checkmark     & 40.6 \\

Star E \\
1      & 6365-56608-154  & $-$78.5 & 
P     & \checkmark     & \checkmark     & 68.2 \\
2      & 6365-56642-165 & $-$78.5 & 
\checkmark     & \checkmark     & \checkmark     & 61.6 \\
3      & 6365-56676-16  & $-$78.2 &
\checkmark       & \checkmark       & \checkmark     & 71.7 \\

Star F\\
1      & 6247-56264-148 & $-$84.8  & 
\checkmark       & \checkmark     & \checkmark     & 165.2 \\
2      & 6247-56283-093 & $-$84.8  & 
\checkmark     & \checkmark     & \checkmark     & 170.5 \\
3      & 6247-56541-099 & $-$85.0  & 
\checkmark       & \checkmark       & \checkmark     & 88.4\\
4      & 6247-56542-099 & $-$85.1  & 
\checkmark     & \checkmark     & \checkmark     & 110.3 \\
	\end{tabular}
    
\end{table*}

\begin{table*}
	\centering
	\contcaption{}
	\begin{threeparttable}

		\begin{tabular}{lcccccl} 
		\hline
Visit $\#$ & Visit ID & Heliocentric Velocity & 
Chip A & Chip B& Chip C & S/N\\
        &  & (kms$^{-1})$ & 
        &         &          &        \\
	\hline

Star G \\
1      & 5282-55822-218  & $-$64.0 
&\checkmark      & \checkmark     & \checkmark     & 7.2 \\
2      & 5282-55823-218  & $-$63.4 & 
\checkmark     & \checkmark     & \checkmark     & 12.9 \\
3      & 5282-55841-212  & $-$63.9 & 
\checkmark      &  \checkmark      & \checkmark     & 9.5 \\
4      & 5283-55816-013  & $-$57.2 &
P     & \checkmark     & \checkmark     & 7.3 \\
5      & 5283-55823-019  & $-$58.4 & 
P     & \checkmark     & \checkmark     & 8.5 \\
6      & 5283-55843-013  & $-$56.9 &
P      & \checkmark     & \checkmark     & 8.2 \\
7      & 5283-55873-013  & $-$61.2 & 
P      & \checkmark     & \checkmark     & 6.4  \\
8      & 5283-55874-013  & 61.7 & 
P      & \checkmark     & \checkmark     & 13.1 \\
9      & 5283-55905-013  & $-$64.9 
&P      &  \checkmark      & \checkmark     & 11.2 \\
10     & 5283-55906-013 & $-$63.2 
&P      & \checkmark     & \checkmark     & 5.4 \\
11     & 6246-56263-013 & $-$63.1 
&P      &  \checkmark      & \checkmark   & 10.4 \\
12     & 6246-56282-019 & $-$68.0 
&P      & \checkmark     & \checkmark     & 5.9 \\
13     & 6246-56318-013 & $-$72.5 
&P      &  \checkmark      & \checkmark     & 5.3 \\
14     & 6246-56539-013 & $-$60.7 
&P     & \checkmark     & \checkmark     & 5.8 \\
15     & 6247-56264-216 & $-$62.4 
&\checkmark      &  \checkmark      & \checkmark     & 11.1 \\
16     & 6247-56283-216 & $-$45.1  
&\checkmark      &  \checkmark      & \checkmark     & 11.7 \\
17     & 6247-56541-211 & $-$60.5  
&\checkmark      &  \checkmark      & \checkmark     & 7.5 \\
18     & 6247-56542-211 & $-$63.1 & 
\checkmark      &  \checkmark      & \checkmark     & 8.6 \\
19     & 6363-56284-013 & $-$68.1 & 
P    &  \checkmark      & \checkmark & 6.4 \\
20     & 6363-56583-014 & $-$63.6 & 
P     & \checkmark     & \checkmark     & 10.6 \\
21     & 6363-56587-013 & $-$62.6 &
P     & \checkmark     & \checkmark     & 9.5 \\
22     & 6364-56285-018 & $-$65.3 &
P     & \checkmark     & \checkmark     & 8.2 \\
23     & 6364-56561-019 & $-$59.9 & 
P     & \checkmark     & \checkmark     & 9.6 \\
24     & 6364-56584-013 & $-$63.2 & 
P     & \checkmark     & \checkmark     & 9.5 \\
25     & 6365-56608-013 & $-$64.1 
&P     & \checkmark     & \checkmark     & 9.1 \\
26     & 6365-56642-018 & $-$60.9  
&P     & \checkmark     & \checkmark     & 8.9 \\
27     & 6365-56676-018 & $-$65.8  
&P     & \checkmark     & O     & 8.7 \\
28     & 6366-56644-013 & $-$62.2  
&P     & \checkmark     & \checkmark     & 8.2 \\

Star H \\
1      & 5282-55815-214  & $-$87.7 
&\checkmark       & \checkmark     & \checkmark     & 6.1 \\
2      & 5282-55822-214 & $-$87.6  
&\checkmark     & \checkmark     & \checkmark     & 26.5 \\
3      & 5282-55823-214  &  $-$87.4 
&\checkmark       & \checkmark       & \checkmark     & 45.5 \\
4      & 5282-55841-232 & $-$87.4 
&\checkmark     & \checkmark     & \checkmark     & 40.4 \\
5      & 5283-55816-016 & $-$87.5 
&P     & \checkmark     & \checkmark     & 25.3 \\
6      & 5283-55823-022  &  $-$87.0 
&P      & \checkmark     & \checkmark     & 33.9 \\
7      & 5283-55843-016  & $-$86.8  
&P      & \checkmark     & \checkmark     & 30.4  \\
8      & 5283-55873-016  & 86.8 & 
\checkmark       & \checkmark     & \checkmark     & 26.7 \\
9      & 5283-55874-016  & $-$87.3 
&P      & \checkmark       & \checkmark     & 47.5 \\
10     & 5283-55905-015  & $-$87.7 
&P      & \checkmark     & \checkmark     & 35.4 \\
11     & 5283-55906-015  & $-$87.5 & 
P      & \checkmark       & \checkmark    & 17.1 \\

		\hline    
    	\end{tabular}
        \begin{tablenotes}
            \item \textit{Note.} P = persistence, F = flat problems, RV = incorrect RV, O = other problems related to SNR or large noise spikes or poor night sky line removal..
        \end{tablenotes}
	\end{threeparttable}

\end{table*}

In order to understand these differences, the individual 
visit spectra for these two stars are examined.  There are 24 visits for Star I and 21 visits 
for Star II, with SNR $>$ 6.
We find a clear persistence problem in many of the spectra, in additional to some other
effects such as poor flat fielding or telluric division problems, poor night sky line
removal, and several cosmic ray hits.

\subsection{Removing Persistence}

Individual visits for Stars I and II were extracted from the APOGEE database.
The alignment of each spectrum was compared to Arcturus, in order to check the radial velocity corrections.
Each visit was then broken into the three wavelength regions corresponding to the blue, green, and red detectors.
Some of APOGEE's detectors suffered from persistence, which is the contamination of a spectrum by remnants of the previous exposure. The persistence problem is worse on the blue chip (1.514-1.581 $\mu$), see Fig. \ref{fig:Figure3}. We remove the portion of the spectrum coming from the blue chip detector for any visit that shows persistence. Occasionally it was also necessary to remove the
green chip spectrum - we suspect that the green chip itself does not have
the persistence problem, but that the data reduction processing of the
visit induces a flat fielding problem when persistence is bad on the blue chip.
After this process, the remaining spectra from each visit are co-added, i.e.,
only the non-persistent spectra from the blue, green, and red regions are kept
for our analysis.

The non-persistent regions of each visit were combined to create the full wavelength range visits, and the cleaned visits were median-combined using IRAF. The final combined spectra for Stars I and II tend
to have fewer green spectra than red, and fewer blue than either.
This results in a lower SNR for the green than red spectrum, and lowest SNR
for the blue spectrum.
These spectra were then normalized with a Legendre polynomial (order=8),
followed by a k-sigma clipping routine (see \citealt{venn2012nucleosynthesis}), and sky lines are removed.  These steps are illustrated in Fig. \ref{fig:Figure3}.
Since these stars are moderately metal-poor, we found this normalization method to be
sufficient for our purposes, but we caution that this is not the same as
that used by the ASPCAP pipeline. Stars G and H also have significant persistence on their spectra.
We have cleaned them similar to Stars I and II.
Stars D, E and F did not have significant persistence problems.
These gave us an opportunity to use and test ASPCAP on the original spectra
in the APOGEE database.

\begin{figure*}
	\includegraphics[width=\columnwidth]{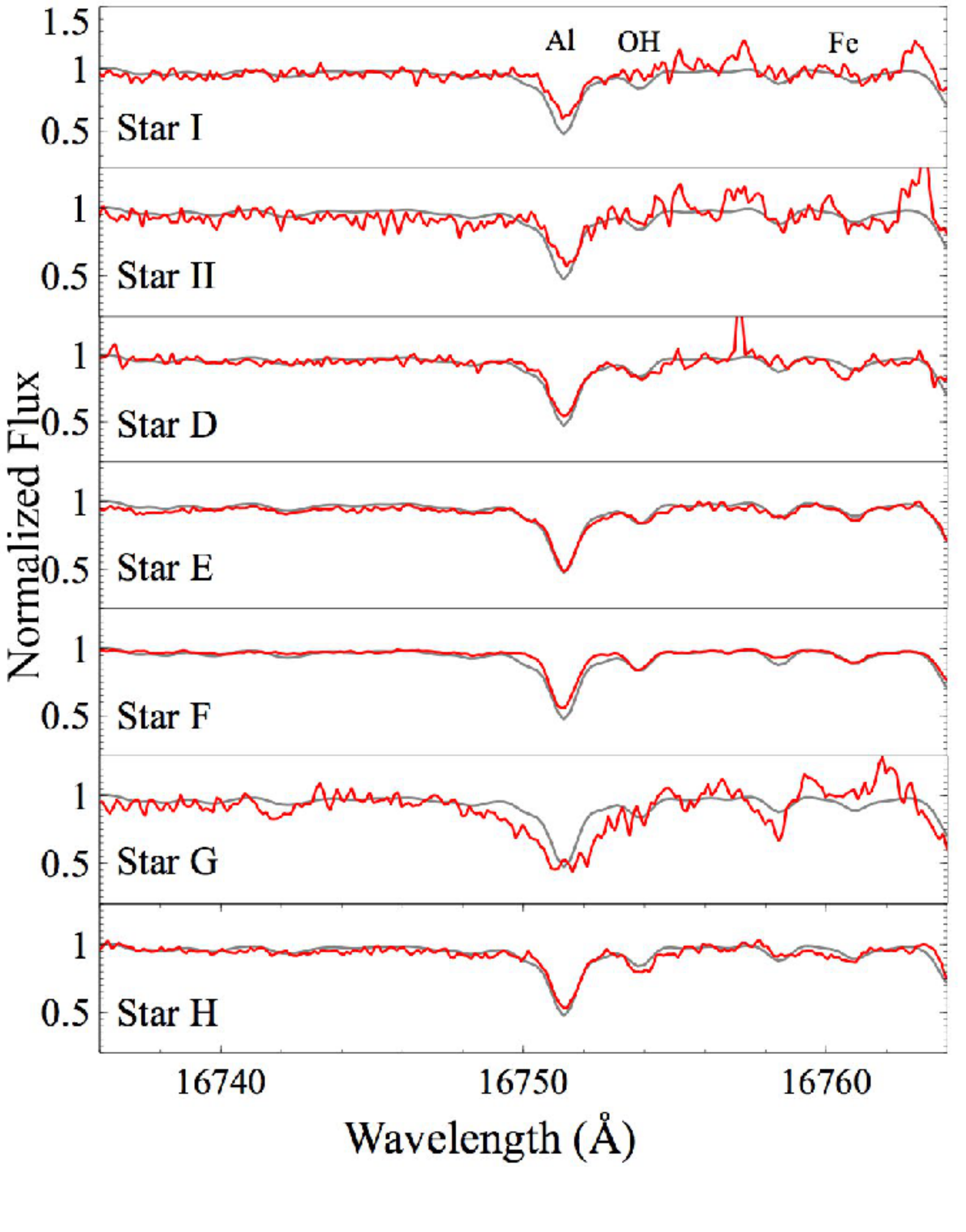}\hfil
    \includegraphics[width=\columnwidth]{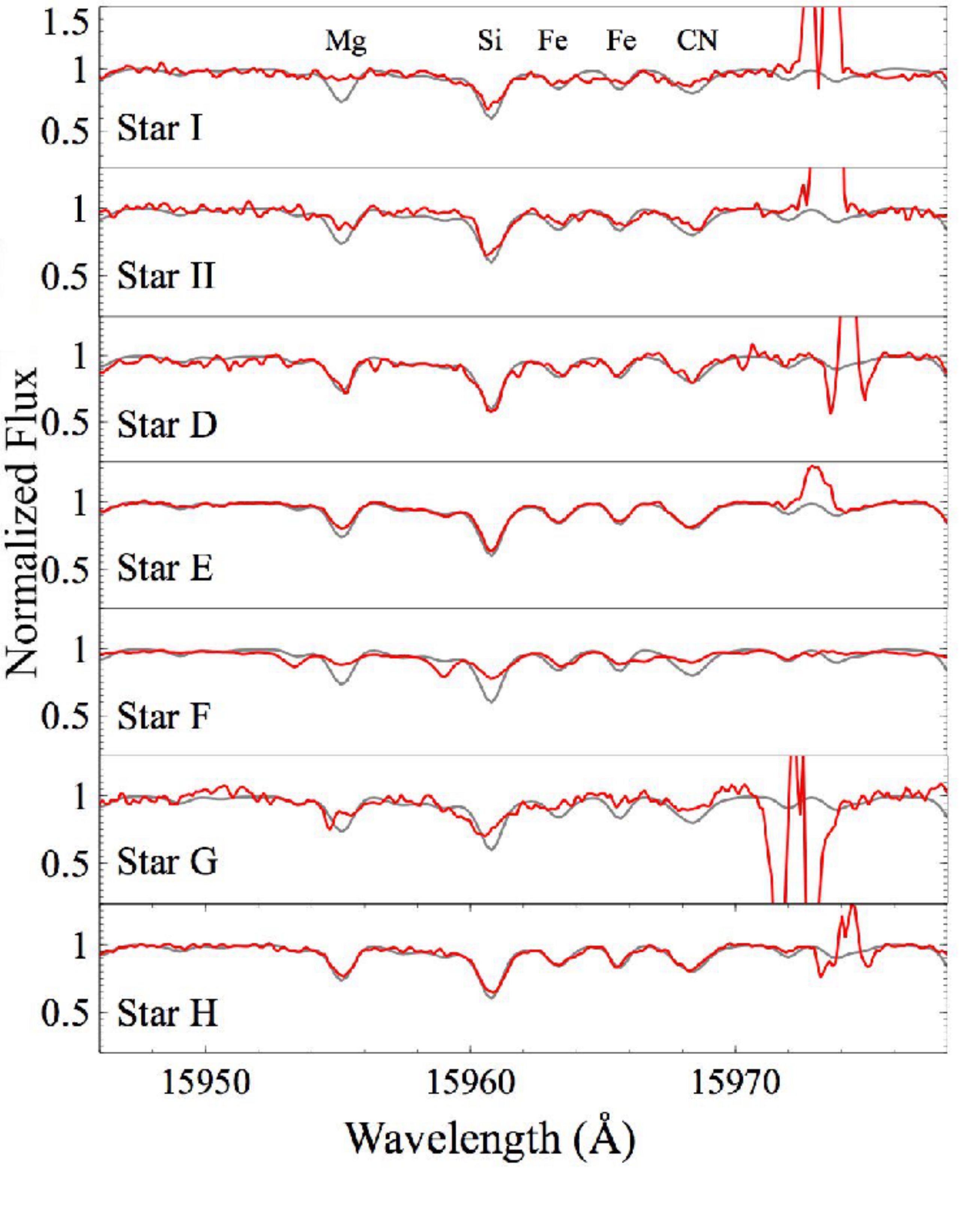}
    \caption{Comparing Pal1 member and candidate spectra (red) to Arcturus (grey).
All spectra have been shifted to the Arcturus wavelength scale.
Lines of Fe~I, Mg~I, Al~I, Si~I, OH and CN are labelled.}
    \label{fig:Figure4}
\end{figure*}

In Fig. \ref{fig:Figure4}, a portion of the cleaned and combined spectra of our Pal1 members to that 
of Arcturus are compared. APOGEE spectra have R=22,500 whereas the Arcturus spectrum 
from (\citealt{hinkle2003atmospheric})  was convolved with a Gaussian profile 
to match the lower resolution and has R=24,000.

Star G shows broader lines than Arcturus and the other
spectra in our sample, which suggests that it is a dwarf star\footnote{ The newest APOGEE DR13 grids for dwarfs include rotation models and therefore log $g$ of Star G is removed in the new data release, which supports our claim that Star G is a dwarf star.}.

In Fig. \ref{fig:Figure4}, the CN, OH, Mg I, Al I, Si I, and Fe I features in our candidate spectra are highlighted and compared to the Arcturus
spectrum. Stars I and II exhibit weaker spectral lines for these species than Arcturus, which can be attributed to their
higher surface temperatures. The aforementioned line broadening observed in Star G is present in these 
spectral ranges as well. 

\section{New Stellar Analyses}

We have carried out a new analysis for all of the stars that may be members of Pal1 based on 
the DR12 data.   This includes those stars that have a persistence problem, but also those
that do not so that we treat the data for all of these objects in a similar way.    
New stellar parameters are determined,  initially from optical and IR
photometry using both the \cite{casagrande2010absolutely} and \cite{ramirez2005effective}, 
colour-temperature relationships.  Temperatures and bolometric corrections are determined from the 
unweighted average of four colours: (B-V), (V-I), (V-K), and (J-K), adopting the metallicity
and cluster distance for Pal1 from \cite{sarajedini2007acs}.
Reddening estimates are from \cite{schlafly2011measuring}.  
Surface gravities are determined photometrically as in \cite{venn2012nucleosynthesis}, 
after adopting a cluster turn-off mass of MA=1.14M$_{\odot}$ (\citealt{sakari2011detailed}) corresponding to its young age, such that:
\begin{equation*}
	logg=4.44+log(MA)+4log(\frac{T_{\rm eff}}{5790}) +0.4(M_{\rm bol}-4.75)
\end{equation*}
\noindent 
The T$_{\rm eff}$ values determined from the two different color-temperature 
calibrations were in excellent agreement for all of the candidates,
with the exception of Star F. 
For this one star, the temperatures differed by $\Delta$T$_{\rm eff} \sim $ 1200 K (see Table \ref{tab:table4}).
The temperature from \cite{casagrande2010absolutely} is much higher, and inconsistent with the position
of this star on the colour-magnitude diagram in Fig. \ref{fig:Figure2}; however, the position of Star F in 
Fig. \ref{fig:Figure2} depends on a correct V magnitude, which has been flagged in the SDSS database.  
Without further information on the V magnitude of Star F, we consider both temperatures
in the discussion below. The difference between the log$g$ values for two different distance moduli from \citeauthor*{harris1996catalog} (1996, 2010 edition) and \cite{sarajedini2007acs} is $\Delta$ log$g$ $\sim 0.4$, which causes only small to negligible differences in our abundance results.

\begin{table}
	\centering
	\caption{Photometric Stellar Parameters }
	\label{tab:table4}
    \begin{threeparttable}
	\begin{tabular}{lrrrrrl} 
		\hline
        & T$_{bv}$ & T$_{vi}$ &T$_{vk}$& T$_{jk}$ & T$_{\rm eff}$& log$g$\\
        & (K)  & (K) &(K)& (K) & (K)& \\ \hline
Star I  & 4930 & 4939    & 4872    & 4481    & 4806 &  2.27 \\
Star II & 5032 & 5015    & 4887    & ---     & 4978 &  2.33 \\
Star D  & 4611 & 4635    & 4582    & 4660    & 4622 & 1.55 \\
Star E  & 4516 & 4554    & 4415    & ---     & 4495 & 1.22  \\
Star F  & 4787 & ---     & ---     & 5199    & 4993 & 1.64  \\
Star F*  & --- &  3800    &  3934    & ---     &  3867 & 0.43  \\ 
Star G  & 4620 & 4632    & 4775    & 4670    & 4674 &  2.28\\ 
Star H  & 4787 & 4805    & 4826    & 4847    &  4816 &  1.47\\
		\hline
	\end{tabular}
        \begin{tablenotes}
            \item \textit{Note.} T$_{\rm eff}$ of Star F* is calculated using \cite{ramirez2005effective} calibration and the rest are computed using \cite{casagrande2010absolutely} calibration.            
        \end{tablenotes}
	\end{threeparttable}
\end{table}

\begin{table*}
	\centering
	\caption{Different Properties of the candidates using FERRE}
	\label{tab:table5}
        	\begin{threeparttable}
	\begin{tabular}{lrrrrr} 
		\hline
 &  DR12 &  DR13 & FERRE\tnote{*}   &  Photom\tnote{*}  &  \cite{sakari2011detailed}
\\ \hline \\
 \bf{Star I}          \\
\\
T$_{\rm eff}$ (K)                            & 5711  	 &  5203             	& 4806   $\pm$ 92       		        &  4806 	$\pm$ 218   & 4800 $\pm$ 70          \\
log$g$                 	           & 3.5    &   2.7             	& 2.8                			   & 2.3 $\pm$ 0.2               	  & 2.27 $\pm$ 0.15       \\
{[}FeI/H{]}                    & $-$0.2     &  $-$0.4  			     & $-$0.8 $\pm$ 0.1         	& $-0.6$ $\pm$ 0.0		& $-0.6$1 $\pm$ 0.08      \\
{[}$\alpha$/Fe{]} 	 & 0.1     	 &   0.2        & ---                 	& 0.1 $\pm$  0.1        & 0.00 $\pm$ 0.00     \\
{[}C/Fe{]}                     & $-$0.2      &   ---        & 0.5 $\pm$ 0.2         & $<$ 0.1        & ---          \\
{[}Ca/Fe{]}                   &  0.4	  & $-$0.4        		 & 0.3 $\pm$ 0.2         		& 0.2 $\pm$ 0.1      	   & 0.16 $\pm$ 0.16    	\\
{[}S/Fe{]}                     &  -0.3      &  ---          & 0.6 $\pm$ 0.2            & 0.2 $\pm$ 0.1         	 & ---        \\
{[}O/Fe{]}                     & 1.0    	 &   0.8     		   & 0.2 $\pm$ 0.4          		&  0.4 $\pm$ 0.1         & \textless0.82     \\
{[}Mg/Fe{]}                   & $-$0.1    &     0.1			 	  & $-$0.3 $\pm$ 0.3      		  & $-$0.1 $\pm$  0.2        & $-$0.11 $\pm$ 0.20     \\
{[}Mn/Fe{]}                   & 0.0	  	&  $-$0.2   			    &  0.9  $\pm$  0.1		 	 & $-$0.1 $\pm$ 0.1         &   ---            \\
{[}Si/Fe{]}                  	 & 0.3	  &   0.1   			     & $-$0.1 $\pm$ 0.1     		   & 0.1 $\pm$ 0.2   	       & 0.24 $\pm$ 0.24      \\
{[}Al/Fe{]}             & ---	  &   ---  			     & 0.1 $\pm$ 0.1  & 0.1 $\pm$ 0.2   	      & ---     \\
{[}K/Fe{]}              & ---	  &   ---  			     & 0.1 $\pm$ 0.1   		   & 0.1 $\pm$ 0.1   	      & ---     \\

\\
 \bf{Star II}          \\
\\

T$_{\rm eff}$ (K)                              &~~~~~ 5602                 & 4886              & 4936  $\pm$ 92      &  4978 $\pm$ 79            & 4750 $\pm$ 135      \\
log$g$                             &~~~~~ 3.2                  & 2.3     	          & 2.7                        & 2.3  $\pm$ 0.2              & 2.33 $\pm$ 0.15      \\
{[}FeI/H{]}                    &~~~~~ $-$0.1                   & $-$0.5       		     & $-$0.5 $\pm$ 0.2      & $-0.6$ $\pm$ 0.0          & $-0.6$1 $\pm$  0.08  \\
{[}$\alpha$/Fe{]}     &~~~~~ 0.1           		         & 0.2   		 & ---                  & 0.1 $\pm$ 0.1                 & $-$0.10 $\pm$ 0.00           \\
{[}C/Fe{]}                     & $-$0.3    							  &   ---        & 0.1 $\pm$ 0.2        					 & $<$ 0.2			   			     & ---          \\
{[}Ca/Fe{]}                   &~~~~~ 0.3                   & $-$0.3          & 0.4 $\pm$ 0.2               & 0.2 $\pm$ 0.3                 & $-$0.04 $\pm$ 0.22	\\
{[}S/Fe{]}                     &~~~~~ $-0.6$                   & ---  & 0.0 $\pm$ 0.2               & 0.2 $\pm$ 0.1                 & ---                    \\
{[}O/Fe{]}                     &~~~~~ $-$0.1                 & 0.7            & 0.1 $\pm$ 0.1 & 0.2 $\pm$ 0.1                 & \textless0.32        \\
{[}Mg/Fe{]}                   &~~~~~ $-0.6$                 & 0.0            &$-$0.2 $\pm$ 0.2              &  $-$0.1 $\pm$ 0.1               & $-$0.13 $\pm$ 0.30       \\
{[}Mn/Fe{]}                   &~~~~~  0.0               &  $-$0.1            &  0.5 $\pm$ 0.1             &  $-$0.1 $\pm$ 0.1               &  $-$0.16 $\pm$ 0.36        \\
 {[}Si/Fe{]}                   &~~~~~ 0.2                   &  0.2           & 0.1 $\pm$ 0.2              & 0.2 $\pm$ 0.1                 & 0.13 $\pm$ 0.23                \\
{[}Al/Fe{]}                  	 & ~~~~~ 0.0	           &    ---   		& $-$0.3 $\pm$  0.1 &  $-$0.1 $\pm$ 0.1   	      &  --- \\
{[}K/Fe{]}                        & $-$0.8 	   						&    ---  			     &    0.0	$\pm$  0.1							   & 0.2 $\pm$ 0.1   	      & ---     \\

       \hline
	\end{tabular}
            \begin{tablenotes}
            \item \textit{*} This use of ``FERRE" is on our persistence cleaned spectra, allowing FERRE to simultaneously determine the stellar parameters and chemical abundances, whereas "Photom" uses our photometrically determined stellar parameters.
        \end{tablenotes}
	\end{threeparttable}
\end{table*}

\begin{table*}
	\centering
	\caption{Different Properties of the candidates using FERRE}
	\label{tab:table6}
	\begin{threeparttable}
    \begin{tabular}{lrrrrrrrr} 
		\hline
 & & DR12 & & DR13 && FERRE\tnote{*}  && Photom\tnote{*} 
\\ \hline \\
 \bf{Star D}          \\
\\
T$_{\rm eff}$ (K)                    &         & 4958     &       &   4843 &      & 4870  $\pm$ 92   &                &  4622 $\pm$ 33   \\
log$g$                    &         & 2.7    &        &   2.6   &       & 2.8  $\pm$ 0.1  &                  & 1.6 $\pm$ 0.2    \\
{[}FeI/H{]}                &   & $-$0.3       &     & $-$0.3	     &     & $-$0.4 $\pm$ 0.1    &   & $-0.6$  $\pm$ 0.0     \\
{[}$\alpha$/Fe{]}      &    & 0.1    &      &   0.1  &  & ---           &      & 0.1 $\pm$  0.1	 \\
{[}C/Fe{]}                   &  & 0.3  &  	      &   ---  &       & 0.3 $\pm$ 0.2     &  					 & 0.1 $\pm$ 0.1	\\
{[}Ca/Fe{]}      &           &  0.3     &        &  0.3     &        & 0.4 $\pm$ 0.1    &       &  0.2 $\pm$ 0.2     \\
{[}S/Fe{]}        &           &  0.1     &         &  ---    &  & 0.1 $\pm$ 0.1          & &  0.1 $\pm$ 0.1  \\
{[}O/Fe{]}         &          & 0.0       &      &  0.5   &            & $-$0.1 $\pm$ 0.4   &       & 0.3 $\pm$ 0.1     \\
{[}Mg/Fe{]}         &        & 0.2         &     &  0.0    &          & 0.2 $\pm$ 0.1      &     & $-$0.2 $\pm$ 0.1      \\
{[}Mn/Fe{]}          &       & $-$0.2         &  &  $-$0.1      &       & 0.4 $\pm$ 0.1         & & $-$0.1 $\pm$  0.1      \\
{[}Si/Fe{]}           &       &  0.2         &    &  0.2	 &           & 0.1 $\pm$ 0.1     &          & 0.1 $\pm$ 0.1         \\
{[}Al/Fe{]}   		   &     & $-$0.7		      &  &    ---   &	    	&  $-0.6$ $\pm$  0.1 &&  0.4 $\pm$ 0.1    \\
{[}K/Fe{]}   		    &    & $-$0.4		       & &     ---   &	    	&  $-$0.2 $\pm$ 0.1     &&  $-$0.2 $\pm$ 0.1    \\
\\
 \bf{Star E}          \\
\\

T$_{\rm eff}$			&					& 4231       &   & 4138     &    & 4168    $\pm$ 92     &       		    & 4495 $\pm$  72  \\
log$g$			&					& 1.5       &   & 1.3      	 &  &  1.9   $\pm$ 0.1       &    		        &  1.2 $\pm$ 0.15   \\

	\end{tabular}
	\end{threeparttable}
\end{table*}

\begin{table*}
	\centering
	\contcaption{}
	\begin{threeparttable}
	\begin{tabular}{lrrrrrrrr} 
		\hline
 &&  DR12 &&  DR13 && FERRE\tnote{*}  &&  Photom\tnote{*} \\ \hline
\\
{[}FeI/H{]} 	&				& $-$0.7         & & $-0.6$   		&  & $-0.6$ $\pm$ 0.1     &         & $-0.6$ $\pm$ 0.0   \\
{[}$\alpha$/Fe{]}	&& 0.1           && 0.1       &      & ---                           & & 0.2 $\pm$ 0.1 \\
{[}C/Fe{]}			 &       & 0.0  &         & ---       &   & $-$0.1 $\pm$ 0.0            & &  0.1 $\pm$ 0.1       \\
{[}Ca/Fe{]}				& 	 & 0.0            && 0.1    	&			  & 0.0 $\pm$ 0.1  &          & 0.2 $\pm$ 0.2  \\
{[}S/Fe{]}				&		&  0.2       &   & ---       &   & 0.1 $\pm$ 0.1      	    &  & 0.2 $\pm$ 0.5  \\
{[}O/Fe{]}				&		& $-$0.2      &    & 0.1     	&		 &  0.1 $\pm$ 0.1        &      & 0.5 $\pm$ 0.1   \\
{[}Mg/Fe{]}				&		& $-$0.3     &     & 0.1     	&		 &  0.0 $\pm$  0.1        &  & 0.3 $\pm$ 0.1  \\
{[}Mn/Fe{]}				&		& 0.0     &      & $-$0.1  	&		   &  0.6 $\pm$  0.0       & & 0.0 $\pm$ 0.2   \\
{[}Si/Fe{]}				&		 & $-$0.2  &        & 0.1      &			& 0.2 $\pm$ 0.1         &   & 0.0 $\pm$ 0.1  \\
{[}Al/Fe{]}   		     &      & 0.2	&	        &    ---  &   &  0.3 $\pm$  0.1          & &  0.3 $\pm$ 0.0    \\
{[}K/Fe{]}   		      &  & 0.0		&     	   &     ---   &	    		 &  0.0 $\pm$  0.1 		&	  &  0.0 $\pm$ 0.1    \\
\\
 \bf{Star G}          \\
\\

T$_{\rm eff}$ (K)		&					& 4564   &      & 4472      &          & 4425   $\pm$ 92   &             & 4674  $\pm$   70  \\
log$g$ 		&					& 3.1     &       &  ---     &       & 4.6   $\pm$ 0.1      &             & 2.3  $\pm$ 0.2               \\
{[}FeI/H{]}  &       		& $-$0.5         &&  0.0          	&		& $-$0.2 $\pm$ 0.1    &   & $-0.6$ $\pm$ 0.0     \\
{[}$\alpha$/Fe{]} && 0.2  &        &  0.1         &   &  ---           &    &  ---   \\
{[}C/Fe{]}  		&	 & 0.5   &       &  ---     &         & 0.2 $\pm$ 0.1   &    & $<$ 0.2  \\
{[}Ca/Fe{]}  		&		& --- & &  0.0        	&		  & 0.0 $\pm$ 0.1   &     & ---      \\
{[}S/Fe{]}   		&		& 0.0  &        &  ---   &           & $-$0.8 $\pm$ 0.2&      & ---      \\
{[}O/Fe{]}  		&		 & 0.3  &        &  0.1   &   			    & 0.1 $\pm$ 0.1   &     & 0.1 $\pm$ 0.1   \\
{[}Mg/Fe{]}			&	  & 0.2      &    & 0.1       &			    & 0.4 $\pm$ 0.1    &    & 0.8 $\pm$ 0.1   \\
{[}Mn/Fe{]} 		&		& $-$0.7    &   & $-$0.1       &			   & 0.2 $\pm$ 0.1   &     & $-$0.1 $\pm$ 0.3   \\
{[}Si/Fe{]}  		&		 & $-$0.0    &    &  0.2      &			    & 0.3 $\pm$ 0.1   &     & 0.2 $\pm$ 0.1  \\
{[}Al/Fe{]}          &       & --- 		&     &    ---   &	         	&  0.0 $\pm$  ---   &      & $<$ 0.7    \\
{[}K/Fe{]}   		  &      & $-$0.1		 &    	   &     --- &  	    				&  0.2   $\pm$ 		0.2			&		  &  $<$ 0.7    \\
\\
\bf{Star H}          \\
\\

T$_{\rm eff}$  (K)                && 4857          &&  4800        && 4780  $\pm$ 92              		      &&  4816 $\pm$  26 \\
log$g$                 && 2.8           &&  2.9          && 3.1  $\pm$ 0.1               		      &&  1.5 $\pm$ 0.2  \\
{[}Fe/H{]}        && $-$0.3            &&  $-$0.3	       && $-$0.3 $\pm$ 0.1    	        && $-$0.4  $\pm$ 0.0  \\
{[}$\alpha$/Fe{]}        &&   0.1          &&   0.1 	       &&  ---     	           && 0.2 $\pm$ 0.1  \\
{[}C/Fe{]}   && 0.2          &&  ---              && $-$0.1 $\pm$ 0.1      && 0.2 $\pm$ 0.1  \\
{[}Ca/Fe{]} 	 &&  0.2      &&  0.1           && 0.0 $\pm$ 0.0      	  	      &&  0.1 $\pm$ 0.0 \\
{[}S/Fe{]}        &&  $-$0.2      &&  ---   && 0.4 $\pm$ 0.1     	  	       &&  0.2 $\pm$ 0.1 \\
{[}O/Fe{]}	       && 0.1              &&  0.3           &&  0.1 $\pm$ 0.2              && 0.2 $\pm$ 0.1 \\
{[}Mg/Fe{]}      && 0.2              &&  0.2           && 0.1 $\pm$ 0.1    	          && 0.5 $\pm$ 0.1 \\
{[}Mn/Fe{]}      && $-$0.2      &&  -0.1         && 0.3 $\pm$ 0.0    	     	     && 0.1 $\pm$  0.2 \\
{[}Si/Fe{]}	        && 0.2             &&   0.1          && 0.1 $\pm$ 0.1                 && 0.1 $\pm$ 0.1  \\
{[}Al/Fe{]}         && 0.0		      &&    ---   && $-$0.2 $\pm$  ---     && 0.3 $\pm$ 0.1    \\
{[}K/Fe{]}         && $-$0.3		      &&    ---   			&& 0.0  $\pm$   0.1 				&& 0.0$<$    \\

       \hline
	\end{tabular}
     \begin{tablenotes}
        \item \textit{*} This use of ``FERRE" is on our persistence cleaned spectra, allowing FERRE to simultaneously determine the stellar parameters and chemical abundances, whereas ``Photom" uses our photometrically determined stellar parameters.
        \end{tablenotes}
	\end{threeparttable}
\end{table*}

\begin{table*}
	\centering
	\caption{Different Properties of Star F}
	\label{tab:table7}
   	\begin{threeparttable}
	\begin{tabular}{lrrrrr} 
		\hline
 &  DR12 &  DR13 & FERRE\tnote{*}  &  Photom1\tnote{*}  & Photom2\tnote{*} \\ \hline \\

 \bf{Star F}          \\
\\
T$_{\rm eff}$ (K) 	        & 4848  & 4828    & 4875 $\pm$ 92      & 4993 $\pm$ 291    & 3867 $\pm$95  \\
log$g$          & 2.6   & 2.5     & 3.2  $\pm$ 0.1     & 1.6  $\pm$ 0.2    &  0.4 $\pm$ 0.2 \\
{[}Fe/H{]}      & $-$0.3  & $-$0.4    & $-$0.8 $\pm$ 0.1     & $-0.6$ $\pm$ 0.1   & $-$1.4 $\pm$0.1 \\
{[}$\alpha$/Fe{]} & 0.1 & 0.1   & ---          & $-$0.3 $\pm$ 0.3    & --- \\
{[}C/Fe{]}        & $-$0.1  & --- & 0.4 $\pm$ 0.3    &  0.3 $\pm$ 0.1    & $-$0.1 $\pm$0.1 \\
	\end{tabular}
    \end{threeparttable}
\end{table*}

\begin{table*}
	\small
	\centering
	\contcaption{}
    \begin{threeparttable}
	\begin{tabular}{lrrrrr} 
		\hline
 &  DR12 &  DR13 & FERRE\tnote{*}  &  Photom1\tnote{*}  & Photom2\tnote{*} \\ \hline \\

{[}Ca/Fe{]}     & $-$0.1  &  0.0      & 0.0 $\pm$ 0.1     & $-$0.3 $\pm$ 0.1   & ---  \\
{[}S/Fe{]}      & 0.2   &  ---  & $-$0.1 $\pm$ 0.2    &  $-$0.4 $\pm$ 0.1  & $-$0.2 $\pm$ 0.1 \\
{[}O/Fe{]}      & 0.1   & 0.1       & $-$0.1 $\pm$ 0.3    & $-$0.2 $\pm$ 0.2   & 0.0 $\pm$ 0.1  \\
{[}Mg/Fe{]}     & 0.0   &  0.1      & 0.2 $\pm$ 0.1     & $-$0.3 $\pm$0.2    & $-$0.5 $\pm$ 0.1 \\
{[}Mn/Fe{]}     & 0.1   & 0.0       &  0.8  $\pm$ 0.0   & 0.3 $\pm$  0.2   & 0.3 $\pm$ 0.1  \\
{[}Si/Fe{]}     & 0.2   & 0.1       &  0.0 $\pm$ 0.0    & $-$0.1 $\pm$ 0.2   & 0.0 $\pm$ 0.1  \\
{[}Al/Fe{]}     & 0.2   &  ---  &  0.6 $\pm$  0.0   & 0.8 $\pm$ 0.1    & 0.8 $\pm$ 0.1 \\
{[}K/Fe{]}      & $-$0.1  &  ---  & $-$0.1  $\pm$ 0.1   & 0.1 $\pm$ 0.1    & $-$0.2 $\pm$0.1 \\ \hline

	\end{tabular}
            \begin{tablenotes}
        \item \textit{*} This use of ``FERRE" is on our persistence cleaned spectra, allowing FERRE to simultaneously determine the stellar parameters and chemical abundances.  For Star F, we found two very different temperatures depending on which set of photometric magnitudes were examined; see Table \ref{tab:table4}.   Here we present the elemental abundances for each temperature.
        \end{tablenotes}
	\end{threeparttable}
\end{table*}

The APOGEE ASPCAP data analysis pipeline uses the least squares template fitting routine,
FERRE (\citealt{prieto2006spectroscopic}), which matches observed spectra to (renormalized) synthetic spectra 
from model atmospheres that have been run through the 1D, LTE, spectrum synthesis code ASSET 
(\citealt{koesterke2008center}; \citealt{koesterke2009quantitative}).  FERRE simultaneously determines the
stellar parameters, metallicities, and element abundance ratios for a given spectrum.
We too have used FERRE\footnote {FERRE at Github: https://github.com/jobovy/apogee.}
for metallicities and chemical abundances, once where FERRE determines the
stellar parameters and a second time where we adopt our photometrically determined 
stellar parameters (see Tables \ref{tab:table5}-\ref{tab:table7}).  
To match the observed spectra to the synthetic spectra, it was necessary to resample
the observations to be on the same wavelength scale.  This caused the observations to
have a slightly lower resolution than the original visits, and the combined spectra had 
a slightly larger spectral range.  This resulted in observations of a few additional absorption 
lines (K, Mn) that that are not in the APOGEE DR12 database. 

For Stars I and II,  Table \ref{tab:table5} shows that the photometric stellar parameters yield 
chemical abundances and metallicities in excellent agreement with the optical analyses. 
This implies that persistence is a significant problem in the analysis of these 
two stars in the DR12 data release
(also see discussion of the DR13 data in Section 6.4).
This further implies that the analysis of some stars in the APOGEE
database can still be improved using the APOGEE spectra themselves.

\section{Stellar Abundances}

The stellar parameters and chemical abundances for 10 elements have been redetermined
in this paper for in a set of Pal1 members and candidates from persistence-cleaned 
APOGEE spectra.    The results are shown in Tables \ref{tab:table5}-\ref{tab:table7}, including the elements C, O, 
Mg, Al, Si, S, K, Ca, Mn, and Fe (see Table \ref{tab:tableA} for log abundances of all detected lines).

The abundance uncertainties are calculated in two ways.
When fewer than four lines are available, the error is taken as the standard deviation in [Fe/H].
When there are more than four lines, the measurement error is taken as the standard deviation 
divided by root of number of lines.   For cases where either of these methods results in an
error $<0.1$ dex, an error of 0.1 dex is adopted since the best synthetic fits have been
determined by eye.

A few elements require special notes:
\begin{itemize}

\item Titanium:
\cite{holtzman2015abundances} show that the APOGEE (DR12) abundances do not reproduce the [Ti/Fe] trends seen for stars in the solar neighbourhood by \cite{bensby2014exploring}.  This difference is not currently understood, and
therefore the ASPCAP titanium lines are to be treated with caution.   \cite{hawkins2016accurate} suggested that the Ti line at 15837.8\AA, which is not included in the set adopted by ASPCAP, can be considered reliable. We did not use this line in our FERRE estimates.

\item $[\alpha$/Fe$]$: 
We estimate a mean [$\alpha$/Fe] ratio by averaging the results for Mg, S, Si and Ca
(not O due to the very noisy oxygen lines, and not Ti as discussed above).   

\end{itemize}

Overall, the chemical abundances of Stars I and II  are in a good agreement with the optical
analysis by \cite{sakari2011detailed}.  Three candidate stars (Stars D, E, and G) have stellar 
parameters typical of red giants and metallicities of [Fe/H]=$-0.6$, when determined from
the photometric parameters.   These values are similar to the members in the core of Pal1.
On the other hand, the chemistry of Star H is sufficiently different that it is a likely non-member. 

Star F warrants special attention due to its position in the tidal tails of Pal1.
Two temperatures have been determined from the color-temperature calibrations for this star,
based on its photometric uncertainties (see Table \ref{tab:table4}).  When the cooler temperature is examined, 
then its metallicity is significantly different from that of Pal1 such that it would be a non-member.  
However, if the  hotter temperature is adopted, its stellar parameters are typical of a red giant, 
with a metallicity and chemical abundances that are similar to those of the members of Pal1.
Furthermore, with the hotter temperature, then Star F has a low [$\alpha$/Fe] that is 
consistent with the other members of Pal1.   Its high [Al/Fe], with slighly low [Mg/Fe],
is unusual for a star in Pal1, unless Star F is, or has been contaminated by, an AGB star
(e.g., \citealt{ventura2008self}).


\section{Discussion}

Using the APOGEE database, we have re-examined the spectra for two known members of Pal1 
and five new candidate members that are well away from the central region of this cluster.
For each member and candidate star, all visits were examined and the blue chips of the spectra with persistence removed, then recombined the clean visits (see section 3.1 for more details). A new stellar analysis has been conducted using FERRE.  The results for the cleaned spectra of Stars I and II are in excellent agreement with the optical analysis by \cite{sakari2011detailed}, whereas the DR12 analyses based on the original spectra are not (see Table \ref{tab:table2}). The chemical abundance and stellar parameters of the candidates are shown in Table \ref{tab:table5}-\ref{tab:table7}. 
The estimated [$\alpha$/Fe] ratios for Stars I and II are  in good agreement with the optical results of \cite{sakari2011detailed}. The Na I lines are too weak or noisy in most of the spectra for reliable determinations of [Na/Fe], therefore we do not investigate the Na-O anticorrelation.

\begin{figure}
	\includegraphics[width=\columnwidth]{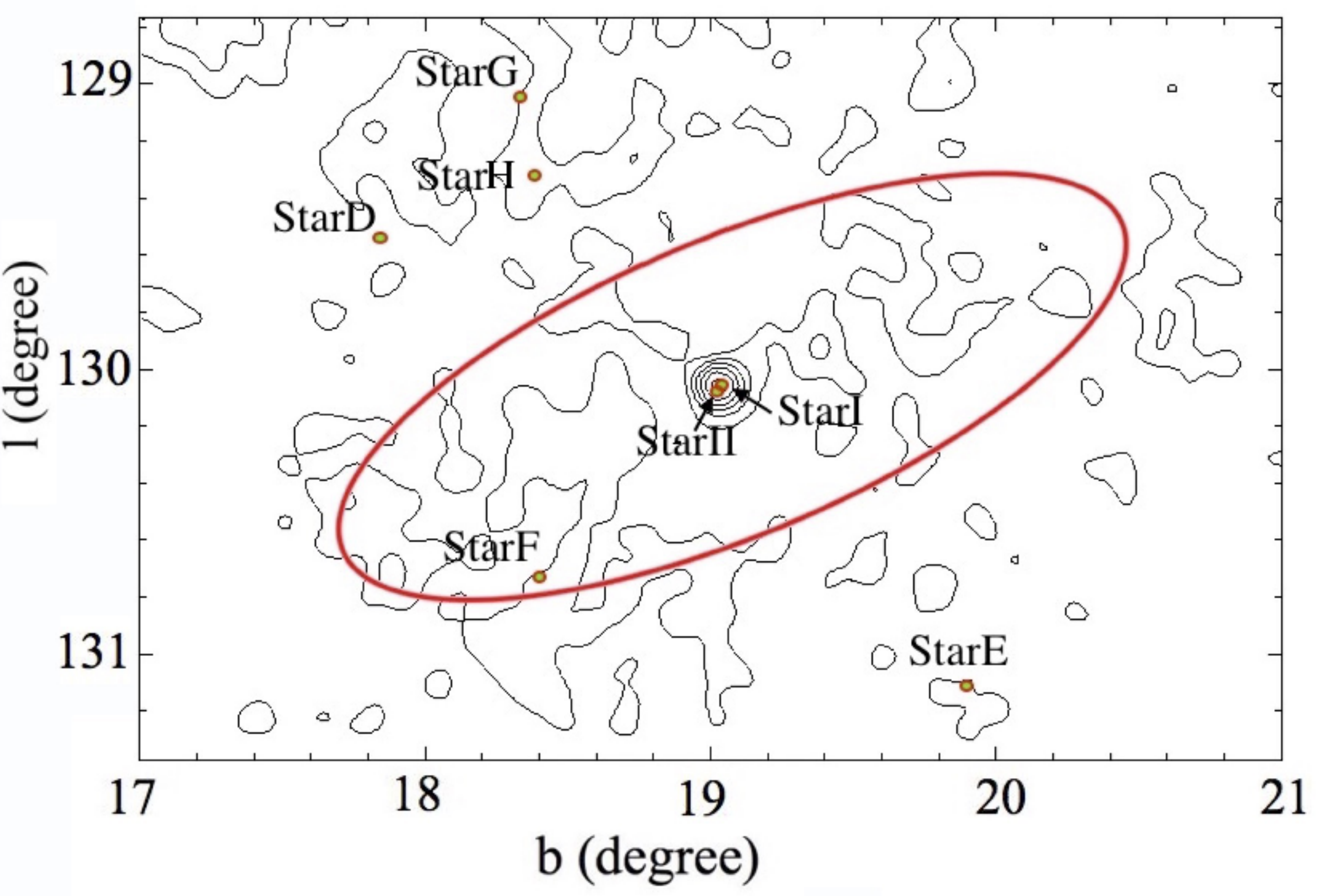}
    \caption{Position diagram of our Pal1 members and candidates relative to the SDSS stellar densities around Pal1. 
		 The red outline shows the position of the tails from Niederste-Ostholt et al. (2010). Our stars do not lie in those tails, except Star F.}
    \label{fig:Figure5}
\end{figure}

\subsection{Tidal Tails of Pal1}

The position of the Pal1 candidates with respect to the tidal tails mapped out
by \cite{niederste2010tidal} based on SDSS photometry are shown in Fig. \ref{fig:Figure5}. Their contour map is constructed from a probability-weighted star count map of Pal1 candidates from the CMD in the MSTO/MS region. The number of candidates per square  arcmin can be determined as 0.856 at the centre of Pal1 and 0.050 above the background region. The positions of Stars I, II, and D-H are shown relative to new isophots determined by M. Irwin from the same SDSS data in Fig. \ref{fig:Figure5}.   A difference in the adopted bin sizes and isophot levels can suppress the apparent tidal features. Only one of these candidates, Star F, is coincident with the tidal tail found by \cite{niederste2010tidal}.

Given the distance to Pal1 as 14.2 kpc from the Sun (\citealt{sarajedini2007acs}), and the angular separations 
of each star from its core, then the minimum distances of each star from the core 
of Pal1 range from 220.8 pc (Star F) and 236.9 pc (Star H), 
to 294.6 pc (Star G), 319.8 pc (Star D), and 326.7 pc (Star E).
For these stars to have reached these distances over
the lifetime of this cluster ($< 6$ Gyr) would have required ejection velocities
$\le$1 kms$^{-1}$.   
These velocities are not particularly large, therefore it is possible that if
stars escape from Pal1 then they could be lurking at these angular separations. 

\begin{figure}
	\includegraphics[width=\columnwidth]{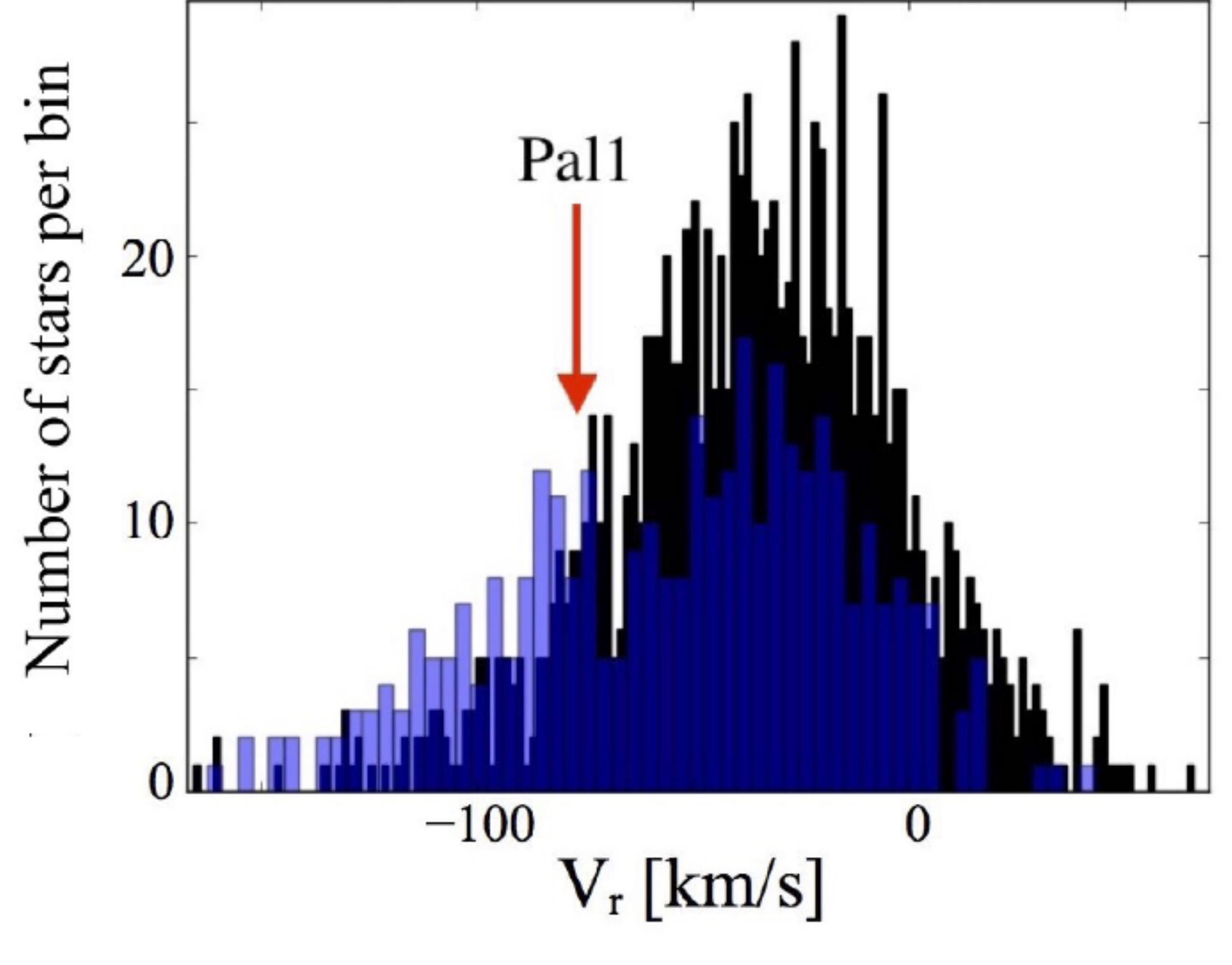}
    \caption{Histogram of the heliocentric radial velocities of APOGEE data 
(blue bars) and as determined from the Besan\c{c}on model (\citealt{robin2003synthetic}) 
in the direction of Pal~1 (black bars, renormalized).  This suggests
that our candidate Pal1 objects could be consistent with sampling of
the smooth Galactic halo distribution.}
    \label{fig:Figure6}
\end{figure}

\subsection{Membership Probability Analysis}

We examined the Besan\c{c}on model (\citealt{robin2003synthetic}) of the stellar populations in the Galaxy to evaluate membership probability of the new Pal1 candidates. The number of stars in the smooth Galactic halo in the direction of Pal1 are estimated based on similar limits in magnitude, colour, radial velocity, and metallicity (Fig. \ref{fig:Figure6}). To extract this simulated dataset, we run 
the model with the following selection criteria:

\begin{itemize}
\item an H-band range of 7 to 13.8, comparable to the APOGEE target list. 
\item a distance interval from 0 to 50 kpc, to include most of both foreground and background stars. 
\item a 7 sq. deg. field of view, centred on Pal1 to match the SDSS field.
\item  The APOGEE database flags all non-giant stars as dwarfs. To directly compare the Besan\c{c}on results with APOGEE, MS, WDs and T Tauri stars were removed from our Besan\c{c}on model and only giant stars were taken into account. 
\end{itemize}

 These selection criteria result in 1124 total stars in the Besan\c{c}on model. 129 (12$\%$)
 stars have radial velocities ($-75$ $\pm$ 15 kms$^{-1}$) and metallicities ([Fe/H] = $-0.6$ $\pm$ 0.4) similar to our parameters for the
 APOGEE search in the Pal1 field. These should be treated as field contaminants from the smooth halo distribution. In comparison,
 the SDSS/APOGEE Pal1 field contains 377 giants, of which 33 (9\%) have radial velocities and metallicities similar to Pal1. Therefore, the Besan\c{c}on model predicts a larger fraction of field contaminants (12\%) than observed
 in the APOGEE Pal1 field (9\%). This strongly implies that the APOGEE field is representative of the smooth halo, with no
 evidence for additional stars due to the Pal1 globular cluster.

It should be noted that APOGEE's Pal1 field is subject to observational placements, particularly in the fibre limitations. These include
 (1) crowding in the centre of Pal1 where the bonafide members are located, (2) that not all red giants can
 be observed simultaneously, and (3) that only 30\% of the total number of good targets were observed.

\begin{figure}
	\includegraphics[width=\columnwidth]{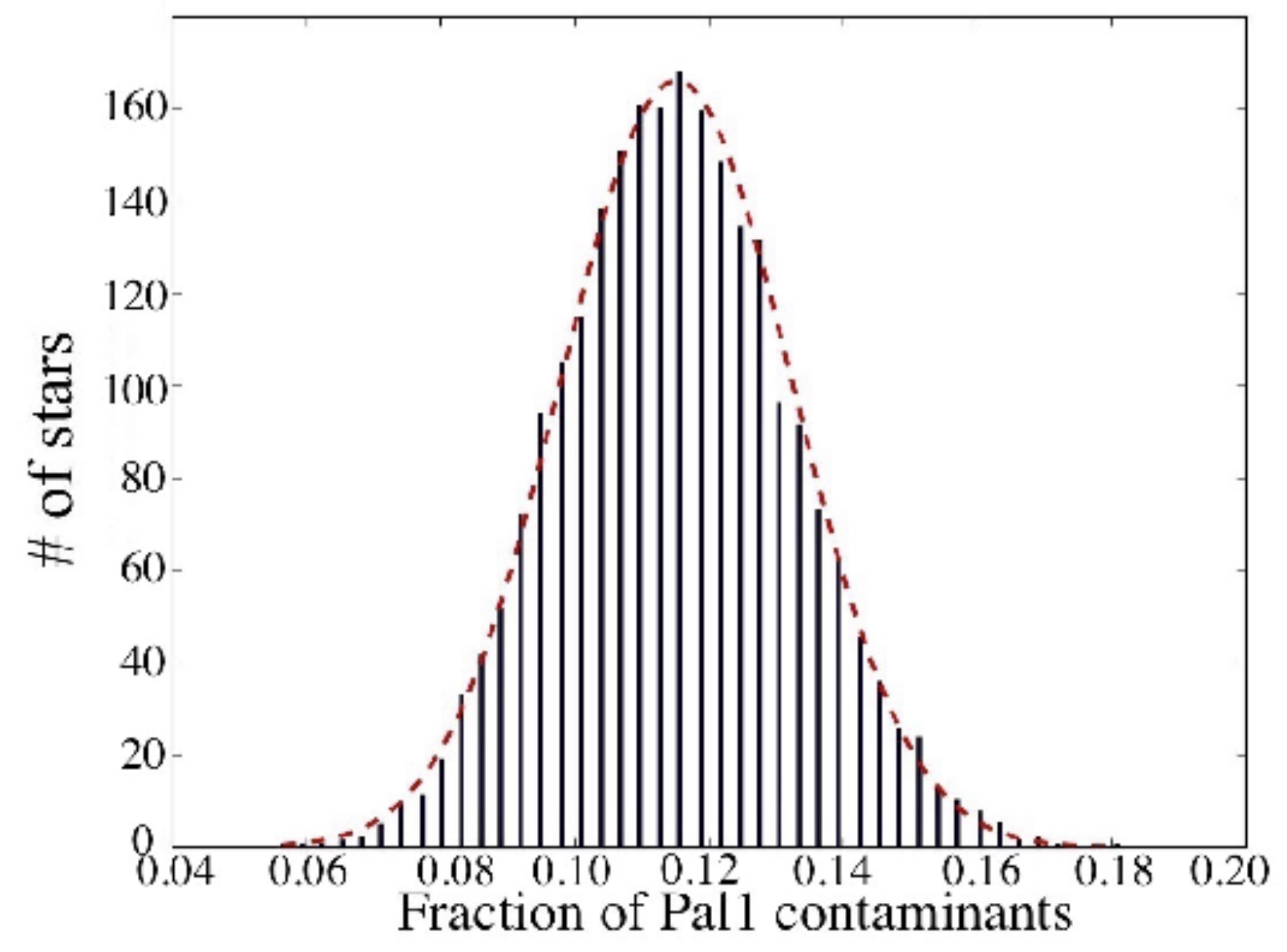}
    \caption{The histogram of the Gaussian distribution of Pal1 field contamination after 10000 runs in the Monte Carlo simulation.
This plot suggests that $\sim$ 12 \% of stars in Pal1 field have radial velocities and metallicities comparable to our search criteria around the Pal1 core.}
    \label{fig:Figure7}
\end{figure}

A Monte Carlo approach was also used to randomly examine the potential for extracting Pal1 members from APOGEE Pal1 field. This was done by selecting 30\% of stars from Besan\c{c}on model to account for the APOGEE selections. For each sampling run, the fraction of field stars with our search criteria for Pal1 radial velocities and metallicities ($-75$ $\pm$ 15 kms$^{-1}$ and [Fe/H] = $-0.6$ $\pm$ 0.4 dex, respectively) was calculated. 10000 runs were performed and the histogram of the distribution of corrected field contaminants is shown in Fig. \ref{fig:Figure7}. This histogram shows a well defined Gaussian distribution with a mean fraction of Pal1 contaminants of ~0.12 $\pm$ 0.02. 

Considering that number of stars in the RV and [Fe/H] search criteria in the APOGEE Pal1 field yielded 33 out of 377 stars (or 9\%), we find that this is consistent with the predicted estimate from our Monte Carlo sampling of the Besan\c{c}on smooth halo.

\begin{figure}
	\includegraphics[width=\columnwidth]{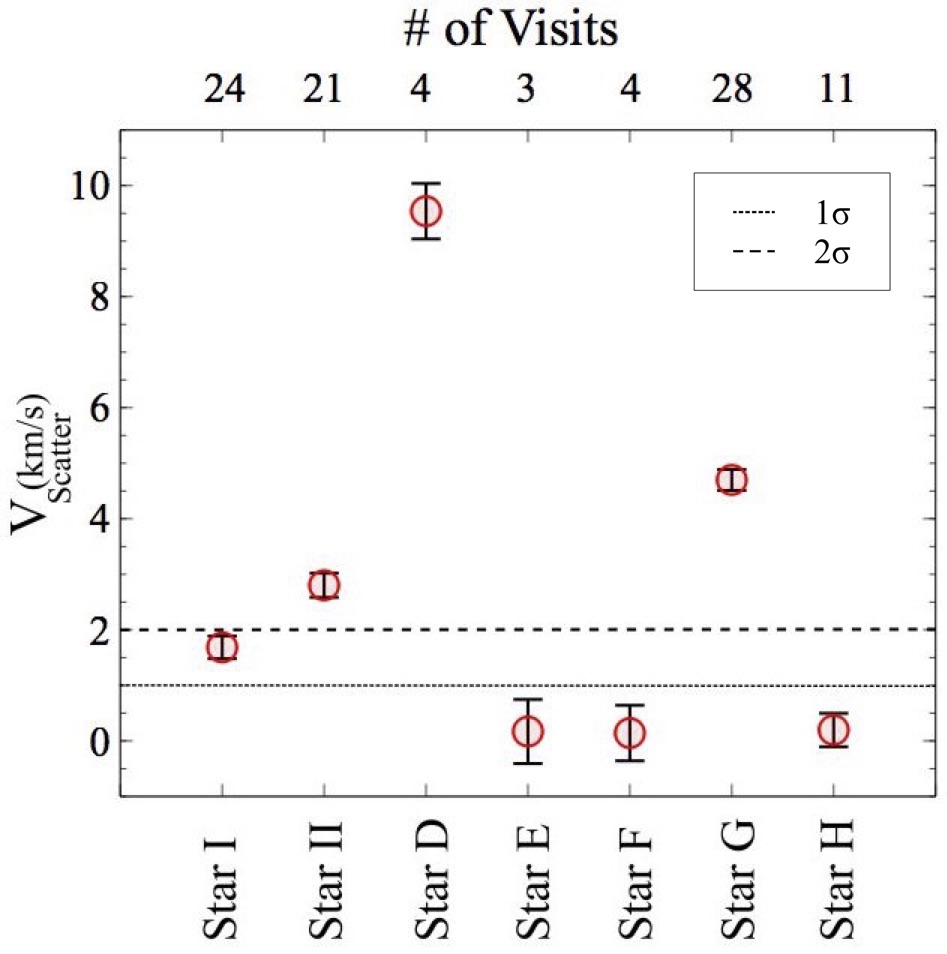}
    \caption{Velocity variation of the Pal1 members and candidates, 
with respect to number of visits for each star. When the scatter is $\ge$ 1 kms$^{-1}$ (dotted line),
there is a high probability of a binary system.   Error bars are 1/$\sqrt{\# ~of ~visits}$.}
    \label{fig:Figure8}
\end{figure}

\subsection{Binarity?}

The velocity variation of the candidates are examined to find any evidence 
for a binarity, which could affect the stellar parameters analysis. 
The radial velocity variations for the two Pal1 members and for all of 
our candidate stars are shown in Fig. \ref{fig:Figure8}. Note that the y-axis represent the RV scatter of the
candidates and is the standard deviation of all visits for each target.
\cite{nidever2015data} has analysed a Plate-to-Plate RV variation analysis
for the APOGEE stars and found that the RVs in the APOGEE database are very stable as the rms scatter is
$\sigma$ = 0.044 kms$^{-1}$. They suggest that stars with RV scatter of greater than 1 kms$^{-1}$ 
have uncertainty much larger than the typical uncertainties and are possibly in a binary system.

Only Stars D and G show a scatter in their radial velocities well above the
2$\sigma$ limit suggested by \cite{nidever2015data} for detecting binary systems.
However, the RV scatter of Star II suggests that it too may be in
a binary system.   If so, the binary nature of this star does not seem to have
affected either its optical analysis, nor our analysis of the corrected IR spectra,
since the stellar parameters and chemical abundance ratios are in good agreement 
with other stars in Pal1. It should be noted that the binarity of Star D, E, F and H cannot be conclusively established as the sample sizes of these stars are small.

\subsection{DR13}

In the SDSS DR13 release, an attempt to unweight spectra with the persistence problem
was established to improve the combined spectral analyses (\citealt{albareti2016thirteenth}). 
When we examine the DR13 database, two more objects could be added to our analysis; however, the results for Stars I and II
are still significantly different from the optical results (see Table \ref{tab:table5}).   
We did not pursue the DR13 data release further.   The persistence problem is indeed well named.

\section{Summary}

Two members of the unusual star cluster Pal1 have been observed in the APOGEE survey.
Examination of their ASPCAP database results are in very poor agreement with previously
determined optical analysis.   We trace this problem to the known persistence problem
that affects up to 30\% of the spectra in the APOGEE database.  By removing those spectra
with persistence (and other reduction problems), we have re-analysed the cleaned spectra.
Our new analyses for the APOGEE spectra of Stars I and II are in excellent agreement with 
the optical analysis by \cite{sakari2011detailed}.  
One star, Star F, may be a member of Pal1, based on its heliocentric radial velocity,
metallicity and chemical abundances, and location in the tidal tails.    However, the
temperature of this star is highly uncertain, and it may be (or be contaminated by) 
an AGB star.  All other candidate members found in the APOGEE DR12 database appear to 
be part of the smooth Galactic background.

\section*{ACKNOWLEDGMENTS}
We would like to thank the anonymous referee for the constructive comments and thorough review of this manuscript.
F.J. and K.A.V. acknowledge funding for this work through the NSERC Discovery Grants program. C.M.S. acknowledges funding from the Kenilworth Fund of the New York Community Trust. 

\bibliographystyle{mnras}
\bibliography{Bibtex}

\begin{thebibliography}{}
\makeatletter
\relax
\def\mn@urlcharsother{\let\do\@makeother \do\$\do\&\do\#\do\^\do\_\do\%\do\~}
\def\mn@doi{\begingroup\mn@urlcharsother \@ifnextchar [ {\mn@doi@}
  {\mn@doi@[]}}
\def\mn@doi@[#1]#2{\def\@tempa{#1}\ifx\@tempa\@empty \href
  {http://dx.doi.org/#2} {doi:#2}\else \href {http://dx.doi.org/#2} {#1}\fi
  \endgroup}
\def\mn@eprint#1#2{\mn@eprint@#1:#2::\@nil}
\def\mn@eprint@arXiv#1{\href {http://arxiv.org/abs/#1} {{\tt arXiv:#1}}}
\def\mn@eprint@dblp#1{\href {http://dblp.uni-trier.de/rec/bibtex/#1.xml}
  {dblp:#1}}
\def\mn@eprint@#1:#2:#3:#4\@nil{\def\@tempa {#1}\def\@tempb {#2}\def\@tempc
  {#3}\ifx \@tempc \@empty \let \@tempc \@tempb \let \@tempb \@tempa \fi \ifx
  \@tempb \@empty \def\@tempb {arXiv}\fi \@ifundefined
  {mn@eprint@\@tempb}{\@tempb:\@tempc}{\expandafter \expandafter \csname
  mn@eprint@\@tempb\endcsname \expandafter{\@tempc}}}

\bibitem[\protect\citeauthoryear{Albareti et~al.,}{Albareti
  et~al.}{2016}]{albareti2016thirteenth}
Albareti F.~D.,  et~al., 2016, arXiv preprint arXiv:1608.02013

\bibitem[\protect\citeauthoryear{Belokurov et~al.,}{Belokurov
  et~al.}{2007}]{belokurov2007orphan}
Belokurov V.,  et~al., 2007, The Astrophysical Journal, 658, 337

\bibitem[\protect\citeauthoryear{Bensby, Feltzing  \& Oey}{Bensby
  et~al.}{2014}]{bensby2014exploring}
Bensby T.,  Feltzing S.,   Oey M.,  2014, Astronomy \& Astrophysics, 562, A71

\bibitem[\protect\citeauthoryear{Bonifacio, Sbordone, Marconi, Pasquini  \&
  Hill}{Bonifacio et~al.}{2004}]{bonifacio2004sgr}
Bonifacio P.,  Sbordone L.,  Marconi G.,  Pasquini L.,   Hill V.,  2004,
  Astronomy \& Astrophysics, 414, 503

\bibitem[\protect\citeauthoryear{Bovy, Erkal  \& Sanders}{Bovy
  et~al.}{2016a}]{bovy2016linear}
Bovy J.,  Erkal D.,   Sanders J.~L.,  2016a, Monthly Notices of the Royal
  Astronomical Society, p. stw3067

\bibitem[\protect\citeauthoryear{Bovy, Bahmanyar, Fritz  \& Kallivayalil}{Bovy
  et~al.}{2016b}]{bovy2016shape}
Bovy J.,  Bahmanyar A.,  Fritz T.~K.,   Kallivayalil N.,  2016b, The
  Astrophysical Journal, 833, 31

\bibitem[\protect\citeauthoryear{Carlberg, Grillmair  \& Hetherington}{Carlberg
  et~al.}{2012}]{carlberg2012pal}
Carlberg R.,  Grillmair C.,   Hetherington N.,  2012, The Astrophysical
  Journal, 760, 75

\bibitem[\protect\citeauthoryear{Carretta et~al.,}{Carretta
  et~al.}{2010}]{carretta2010detailed}
Carretta E.,  et~al., 2010, Astronomy \& Astrophysics, 520, A95

\bibitem[\protect\citeauthoryear{Casagrande, Ram{\'\i}rez, Melendez, Bessell
  \& Asplund}{Casagrande et~al.}{2010}]{casagrande2010absolutely}
Casagrande L.,  Ram{\'\i}rez I.,  Melendez J.,  Bessell M.,   Asplund M.,
  2010, Astronomy \& Astrophysics, 512, A54

\bibitem[\protect\citeauthoryear{Chou, Majewski, Cunha, Smith, Patterson  \&
  Mart{\'\i}nez-Delgado}{Chou et~al.}{2010}]{chou2010chemical}
Chou M.-Y.,  Majewski S.~R.,  Cunha K.,  Smith V.~V.,  Patterson R.~J.,
  Mart{\'\i}nez-Delgado D.,  2010, The Astrophysical Journal Letters, 720, L5

\bibitem[\protect\citeauthoryear{Cohen}{Cohen}{2004}]{cohen2004palomar}
Cohen J.~G.,  2004, The Astronomical Journal, 127, 1545

\bibitem[\protect\citeauthoryear{Doi et~al.,}{Doi
  et~al.}{2010}]{doi2010photometric}
Doi M.,  et~al., 2010, The Astronomical Journal, 139, 1628

\bibitem[\protect\citeauthoryear{Dotter, Chaboyer, Jevremovi{\'c}, Kostov,
  Baron  \& Ferguson}{Dotter et~al.}{2008}]{dotter2008dartmouth}
Dotter A.,  Chaboyer B.,  Jevremovi{\'c} D.,  Kostov V.,  Baron E.,   Ferguson
  J.~W.,  2008, The Astrophysical Journal Supplement Series, 178, 89

\bibitem[\protect\citeauthoryear{Frebel \& Norris}{Frebel \&
  Norris}{2015}]{frebel2015near}
Frebel A.,  Norris J.~E.,  2015, Annual Review of Astronomy and Astrophysics,
  53, 631

\bibitem[\protect\citeauthoryear{Grillmair}{Grillmair}{2006}]{grillmair2006detection}
Grillmair C.,  2006, The Astrophysical Journal Letters, 645, L37

\bibitem[\protect\citeauthoryear{Gunn et~al.,}{Gunn et~al.}{2006}]{gunn20062}
Gunn J.~E.,  et~al., 2006, The Astronomical Journal, 131, 2332

\bibitem[\protect\citeauthoryear{Harris}{Harris}{1996}]{harris1996catalog}
Harris W.~E.,  1996, The Astronomical Journal, 112, 1487

\bibitem[\protect\citeauthoryear{Hawkins, Masseron, Jofre, Gilmore, Elsworth
  \& Hekker}{Hawkins et~al.}{2016}]{hawkins2016accurate}
Hawkins K.,  Masseron T.,  Jofre P.,  Gilmore G.,  Elsworth Y.,   Hekker S.,
  2016, Astronomy \& Astrophysics, 594, A43

\bibitem[\protect\citeauthoryear{Hill, Fran{\c{c}}ois, Spite, Primas  \&
  Spite}{Hill et~al.}{2000}]{hill2000age}
Hill V.,  Fran{\c{c}}ois P.,  Spite M.,  Primas F.,   Spite F.,  2000, arXiv
  preprint astro-ph/0009273

\bibitem[\protect\citeauthoryear{Hinkle, Wallace  \& Livingston}{Hinkle
  et~al.}{2003}]{hinkle2003atmospheric}
Hinkle K.,  Wallace L.,   Livingston W.,  2003, in Bulletin of the American
  Astronomical Society. p.~1260

\bibitem[\protect\citeauthoryear{Holtzman et~al.,}{Holtzman
  et~al.}{2015}]{holtzman2015abundances}
Holtzman J.~A.,  et~al., 2015, The Astronomical Journal, 150, 148

\bibitem[\protect\citeauthoryear{Ishigaki, Hwang, Chiba  \& Aoki}{Ishigaki
  et~al.}{2016}]{ishigaki2016line}
Ishigaki M.~N.,  Hwang N.,  Chiba M.,   Aoki W.,  2016, The Astrophysical
  Journal, 823, 157

\bibitem[\protect\citeauthoryear{Jordi, Grebel  \& Ammon}{Jordi
  et~al.}{2006}]{jordi2006empirical}
Jordi K.,  Grebel E.~K.,   Ammon K.,  2006, Astronomy \& Astrophysics, 460, 339

\bibitem[\protect\citeauthoryear{Koesterke, Prieto  \& Lambert}{Koesterke
  et~al.}{2008}]{koesterke2008center}
Koesterke L.,  Prieto C.~A.,   Lambert D.~L.,  2008, The Astrophysical Journal,
  680, 764

\bibitem[\protect\citeauthoryear{Koesterke, Hubeny, Stone, MacGregor  \&
  Werner}{Koesterke et~al.}{2009}]{koesterke2009quantitative}
Koesterke L.,  Hubeny I.,  Stone J.~M.,  MacGregor K.,   Werner K.,  2009, in
  AIP Conference Proceedings. pp 73--84

\bibitem[\protect\citeauthoryear{Majewski, Schiavon, Frinchaboy  \& et
  al.}{Majewski et~al.}{2015}]{Majewski2015}
Majewski S.,  Schiavon R.,  Frinchaboy P.,   et al. 2015, arXiv preprint
  arXiv:1701.03802

\bibitem[\protect\citeauthoryear{Martin, Ibata, Conn, Lewis, Bellazzini, Irwin
  \& McConnachie}{Martin et~al.}{2004}]{martin2004canis}
Martin N.,  Ibata R.,  Conn B.,  Lewis G.,  Bellazzini M.,  Irwin M.,
  McConnachie A.,  2004, Monthly Notices of the Royal Astronomical Society,
  355, L33

\bibitem[\protect\citeauthoryear{Monaco, Saviane, Correnti, Bonifacio  \&
  Geisler}{Monaco et~al.}{2011}]{monaco2011high}
Monaco L.,  Saviane I.,  Correnti M.,  Bonifacio P.,   Geisler D.,  2011,
  Astronomy \& Astrophysics, 525, A124

\bibitem[\protect\citeauthoryear{Mucciarelli, Carretta, Origlia  \&
  Ferraro}{Mucciarelli et~al.}{2008}]{mucciarelli2008chemical}
Mucciarelli A.,  Carretta E.,  Origlia L.,   Ferraro F.~R.,  2008, The
  Astronomical Journal, 136, 375

\bibitem[\protect\citeauthoryear{Nidever et~al.,}{Nidever
  et~al.}{2015}]{nidever2015data}
Nidever D.~L.,  et~al., 2015, The Astronomical Journal, 150, 173

\bibitem[\protect\citeauthoryear{Niederste-Ostholt, Belokurov, Evans, Koposov,
  Gieles  \& Irwin}{Niederste-Ostholt et~al.}{2010}]{niederste2010tidal}
Niederste-Ostholt M.,  Belokurov V.,  Evans N.,  Koposov S.,  Gieles M.,
  Irwin M.,  2010, Monthly Notices of the Royal Astronomical Society: Letters,
  408, L66

\bibitem[\protect\citeauthoryear{P{\'e}rez et~al.,}{P{\'e}rez
  et~al.}{2016}]{perez2016aspcap}
P{\'e}rez A. E.~G.,  et~al., 2016, The Astronomical Journal, 151, 144

\bibitem[\protect\citeauthoryear{Prieto, Beers, Wilhelm, Newberg, Rockosi,
  Yanny  \& Lee}{Prieto et~al.}{2006}]{prieto2006spectroscopic}
Prieto C.~A.,  Beers T.~C.,  Wilhelm R.,  Newberg H.~J.,  Rockosi C.~M.,  Yanny
  B.,   Lee Y.~S.,  2006, The Astrophysical Journal, 636, 804

\bibitem[\protect\citeauthoryear{Ram{\'\i}rez \& Mel{\'e}ndez}{Ram{\'\i}rez \&
  Mel{\'e}ndez}{2005}]{ramirez2005effective}
Ram{\'\i}rez I.,  Mel{\'e}ndez J.,  2005, The Astrophysical Journal, 626, 465

\bibitem[\protect\citeauthoryear{Robin, Reyl{\'e}, Derri{\`e}re  \&
  Picaud}{Robin et~al.}{2003}]{robin2003synthetic}
Robin A.~C.,  Reyl{\'e} C.,  Derri{\`e}re S.,   Picaud S.,  2003, Astronomy \&
  Astrophysics, 409, 523

\bibitem[\protect\citeauthoryear{Sakari, Venn, Irwin, Aoki, Arimoto  \&
  Dotter}{Sakari et~al.}{2011}]{sakari2011detailed}
Sakari C.~M.,  Venn K.~A.,  Irwin M.,  Aoki W.,  Arimoto N.,   Dotter A.,
  2011, The Astrophysical Journal, 740, 106

\bibitem[\protect\citeauthoryear{Sakari, McWilliam  \& Wallerstein}{Sakari
  et~al.}{2017}]{sakari2017chemical}
Sakari C.,  McWilliam A.,   Wallerstein G.,  2017, arXiv preprint
  arXiv:1701.03802

\bibitem[\protect\citeauthoryear{Salaris \& Weiss}{Salaris \&
  Weiss}{2002}]{salaris2002homogeneous}
Salaris M.,  Weiss A.,  2002, Astronomy \& Astrophysics, 388, 492

\bibitem[\protect\citeauthoryear{Sarajedini et~al.,}{Sarajedini
  et~al.}{2007}]{sarajedini2007acs}
Sarajedini A.,  et~al., 2007, The Astronomical Journal, 133, 1658

\bibitem[\protect\citeauthoryear{Sbordone, Bonifacio, Buonanno, Marconi, Monaco
   \& Zaggia}{Sbordone et~al.}{2007}]{sbordone2007exotic}
Sbordone L.,  Bonifacio P.,  Buonanno R.,  Marconi G.,  Monaco L.,   Zaggia S.,
   2007, Astronomy \& Astrophysics, 465, 815

\bibitem[\protect\citeauthoryear{Schlafly \& Finkbeiner}{Schlafly \&
  Finkbeiner}{2011}]{schlafly2011measuring}
Schlafly E.~F.,  Finkbeiner D.~P.,  2011, The Astrophysical Journal, 737, 103

\bibitem[\protect\citeauthoryear{Siegel et~al.,}{Siegel
  et~al.}{2007}]{siegel2007acs}
Siegel M.~H.,  et~al., 2007, The Astrophysical Journal Letters, 667, L57

\bibitem[\protect\citeauthoryear{Tolstoy, Hill  \& Tosi}{Tolstoy
  et~al.}{2009}]{tolstoy2009star}
Tolstoy E.,  Hill V.,   Tosi M.,  2009, Annual Review of Astronomy and
  Astrophysics, 47, 371

\bibitem[\protect\citeauthoryear{Venn, Irwin, Shetrone, Tout, Hill  \&
  Tolstoy}{Venn et~al.}{2004}]{venn2004stellar}
Venn K.~A.,  Irwin M.,  Shetrone M.~D.,  Tout C.~A.,  Hill V.,   Tolstoy E.,
  2004, The Astronomical Journal, 128, 1177

\bibitem[\protect\citeauthoryear{Venn et~al.,}{Venn
  et~al.}{2012}]{venn2012nucleosynthesis}
Venn K.~A.,  et~al., 2012, The Astrophysical Journal, 751, 102

\bibitem[\protect\citeauthoryear{Ventura \& D'Antona}{Ventura \&
  D'Antona}{2008}]{ventura2008self}
Ventura P.,  D'Antona F.,  2008, Astronomy \& Astrophysics, 479, 805

\bibitem[\protect\citeauthoryear{Villanova, Geisler, Carraro, Bidin  \&
  Mu{\~n}oz}{Villanova et~al.}{2013}]{villanova2013ruprecht}
Villanova S.,  Geisler D.,  Carraro G.,  Bidin C.~M.,   Mu{\~n}oz C.,  2013,
  The Astrophysical Journal, 778, 186

\bibitem[\protect\citeauthoryear{Zasowski et~al.,}{Zasowski
  et~al.}{2013}]{zasowski2013target}
Zasowski G.,  et~al., 2013, The Astronomical Journal, 146, 81

\makeatother
\end{thebibliography}

\appendix

\section{Spectral lines and abundances}

\begin{table*}
	\centering
	\caption{Atomic line data and FERRE [X/Fe]$^a$ ratios}
	\label{tab:tableA}
    \begin{threeparttable}
	\begin{tabular}{llllllllll} 
		\hline
	Element & Lambda (\AA)  & Star I   & Star II  & Star D  & Star E& Star F & Star F* & Star G & Star H  \\ \hline
Fe I             &    15211.682           & $-$1.0               & $-$0.6             & ---                & $-$0.7             & $-$1.0               & \textless$-$1.7   & $-$0.6             & $-$0.4          \\
&15249.140                                & $-$0.6             & $-$0.6             & $-$1.0               & $-$0.7             & $-$0.6             & $-$1.3            & $-$0.4             & $-$0.4          \\
&15297.317                               & $-$0.4             & \textgreater$-$0.4 & \textgreater$-$1.0   & $-$0.7             & ---                & ---               & ---                & $-$0.2          \\
&15309.789                               & ---                & ---                & ---                & $\sim$-0.8       & $-$0.4             & ---               & ---                & $-$0.2          \\
&15339.574                               & $-$0.8             & $-$0.6             & $-$0.9             & $-$0.6             & $-$1.0               & \textless$-$1.8   & $-$0.4             & $-$0.2          \\
&15392.011                               & $-$0.8             & $-$0.6             & $-$0.8             & $-$0.6             & $-$0.6             & \textless$-$1.8   & $-$0.8             & $-$0.6          \\
&15483.107                               & $-$0.6             & $-$0.4             & ---                & $-$0.6             & ---                & ---               & ---                & $-$0.2          \\
&15494.762                               & $-$0.6             & $-$0.6             & $-$0.6             & $-$0.6             & $-$0.6             & ---               & ---                & $-$0.4          \\
&15505.316                               & ---                & ---                & $-$0.6             & $-$0.6             & $-$0.6             & $-$1.4            & $-$0.6             & $-$0.4          \\
&15528.553                               & ---                & ---                & $-$0.4             & $-$0.7             & ---                & ---               & ---                & ---             \\
&15546.326                               & ---                & ---                & \textgreater$-$0.9 & $-$0.5             & $-$0.6             & $-$1.4            & $-$0.4             & ---             \\
&15595.760                                & $-$0.8             & $-$0.6             & $-$1.0               & $-$0.7             & \textless $-$0.8   & $-$1.6            & ---                & $-$0.3          \\
&15608.487                               & $-$0.8             & $-$0.8             & \textgreater$-$1.0   & $-$0.8             & $-$0.6             & $-$1.6            & $-$0.6             & $-$0.4          \\
&15615.412                               & ---                & $-$0.6             & ---                & ---                & ---                & \textless$-$1.8   & $-$0.6             & $-$0.6          \\
&15666.296                               & $-$0.8             & $-$0.8             & ---                & $-$0.7             & \textless $-$0.7   & \textless$-$1.8   & $-$0.5             & $-$0.6          \\
&15681.805                               & $-$0.7             & $-$0.4             & $-$1.0               & $-$0.6             & $-$0.6             & ---               & $-$0.5             & $-$0.4          \\
&15735.713                               & ---                & ---                & ---                & ---                & ---                & ---               & ---                & ---             \\
&15765.622                               & ---                & ---                & ---                & $-$0.4             & $-$0.6             & ---               & ---                & $-$0.2          \\
&15778.381                               & $-$0.4             & $-$0.6             & $-$1.0               & $-$0.5             & $-$0.6             & $-$1.2            & ---                & $-$0.4          \\
&15899.571                               & $-$0.6             & ---                & $-$0.4             & \textgreater$-$0.4 & ---                & ---               & $-$0.5             & $-$0.4          \\
&15905.797                               & $-$0.6             & ---                & $-$0.4             & $-$0.4             & ---                & ---               & ---                & $-$0.4          \\
&15910.390                                & $-$0.6             & $-$0.4             & $-$0.4             & $-$0.5             & $-$0.8             & ---               & ---                & $-$0.4          \\
&15924.987                               & $-$0.7             & $-$0.5             & $-$0.4             & $-$0.7             & $-$0.6             & $-$1.3            & ---                & $-$0.4          \\
&15946.207                               & ---                & ---                & $-$0.4             & $-$0.8             & $-$0.6             & $-$1.6            & ---                & $-$0.5          \\
&15958.447                               & \textless$-$0.8    & $-$0.6             & $-$0.4             & $-$0.6             & ---                & \textless$-$1.4   & $-$0.7             & $-$0.2          \\
&15969.209                               & $-$0.8             & $-$0.8             & $-$0.6             & \textless$-$0.8    & ---                & ---               & $-$0.8             & $-$0.6          \\
&15975.615                               & ---                & ---                & ---                & ---                & $-$0.4             & ---               & ---                & ---             \\
&16011.133                               & $-$0.6             & $-$0.6             & $-$0.4             & $-$0.8             & ---                & $-$1.4            & \textgreater$-$0.4 & $-$0.6          \\
&16013.985                               & $-$0.6             & $-$0.6             & $-$0.4             & $-$0.8             & ---                & ---               & \textgreater$-$0.4 & $-$0.6          \\
&16080.311                               & $-$0.5             & $-$0.6             & ---                & $-$0.6             & $-$0.6             & \textless$-$1.8   & ---                & $-$0.6          \\
&16130.274                               & $-$0.8             & $-$0.6             & ---                & $-$0.6             & $-$0.8             & $-$1.4            & ---                & $-$0.4          \\
&16157.660                                & $-$0.6             & $-$0.4             & $-$0.4             & $-$0.6             & $-$0.6             & $-$1.4            & $-$0.4             & $-$0.4          \\
&16169.448                               & $-$0.6             & $-$0.5             & $-$0.8             & $-$0.7             & $-$0.8             & $-$1.6            & \textgreater$-$0.4 & $-$0.2          \\
&16190.224                               & $-$0.6             & $-$0.4             & $-$1.0               & $-$0.8             & $-$0.6             & ---               & ---                & $-$0.6          \\
&16212.175                               & $-$0.7             & $-$0.4             & $-$0.8             & $-$0.6             & $-$0.8             & ---               & $-$0.8             & $-$0.4          \\
&16217.970                                & $-$0.4             & $-$0.4             & $-$0.8             & $-$0.7             & ---                & $-$1.2            & $-$0.4             & $-$0.4          \\
&16236.084                               & $-$0.8             & $-$0.6             & ---                & $-$0.7             & $-$0.8             & \textless$-$1.8   & ---                & $-$0.6          \\
&16240.487                               & ---                & ---                & ---                & ---                & ---                & \textless$-$1.8   & $-$0.4             & ---             \\
&16256.993                               & ---                & ---                & $-$0.4             & \textless$-$0.8    & $-$0.6             & ---               & ---                & ---             \\
&16297.294                               & $-$0.6             & ---                & ---                & $-$0.8             & $-$0.6             & $-$1.4            & $-$0.4             & $-$0.6          \\
&16320.829                               & ---                & ---                & $-$0.8             & $-$0.6             & $-$0.8             & $-$1.5            & ---                & $-$0.6          \\
&16328.912                               & $-$0.7             & $-$0.4             & $-$0.4             & $-$0.8             & $-$0.8             & \textless$-$1.4   & ---                & $-$0.6          \\
&16402.650                                & $-$0.6             & $-$0.6             & $-$0.8             & $-$0.8             & ---                & \textless$-$1.8   & $-$0.6             & $-$0.6          \\
&16409.869                               & ---                & ---                & ---                & ---                & ---                & ---               & ---                & ---             \\
&16510.805                               & $-$0.6             & $-$0.4             & $-$0.8             & $-$0.6             & ---                & $-$1.4            & $-$0.4             & $-$0.6          \\
&16521.738                               & $-$0.5             & $-$0.6             & $-$0.4             & $-$0.6             & $-$0.4             & \textgreater1.4 & \textgreater$-$0.4 & $-$0.2          \\
&16536.502                               & $-$0.6             & ---                & $-$0.6             & $-$0.6             & $-$0.4             & $-$1.4            & $-$0.6             & $-$0.4          \\
&16556.519                               & $-$0.6             & $-$0.5             & $-$0.4             & $-$0.6             & $-$0.3             & ---               & ---                & $-$0.4          \\
&16590.582                               & ---                & ---                & ---                & ---                & ---                & ---               & ---                & ---             \\
&16617.302                               & ---                & ---                & \textgreater$-$1.0   & \textless$-$0.8    & ---                & \textless$-$1.8   & $-$0.6             & \textless$-$0.6 \\
&16624.278                               & ---                & ---                & $-$1.0               & $-$0.7             & ---                & ---               & ---                & $-$0.6          \\
&16650.424                               & $-$0.4             & $-$0.4             & $-$0.6             & $-$0.6             & $-$0.3             & $-$1.4            & \textgreater$-$0.4 & $-$0.4          \\
&16670.037                               & \textgreater$-$0.4 & \textgreater$-$0.4 & $-$0.8             & $-$0.3             & $-$0.3             & $-$1.4            & ---                & $-$0.4          \\
&16757.644                               & ---                & $-$0.8             & $-$0.4             & $-$0.6             & $-$0.3             & ---               & ---                & $-$0.2          \\
&16804.240                                & ---                & ---                & $-$0.4             & $-$0.6             & ---                & ---               & ---                & $-$0.4          \\
&16848.118                               & ---                & ---                & $-$0.8             & \textless$-$0.8    & ---                & \textless$-$1.8   & ---                & ---             \\
&                  &                  &                  &                  &                  &                 &                  &               \\
O from OH lines   &           15241.164 & ---                & ---                & ---                & 0.0                & ---                & ---               & ---                & 0.2           \\
&15396.206                               & ---                & ---                & ---                & ---                & $-$0.2             & ---               & ---                & 0.4           \\
&15413.211                               & 0.4              & 0.2              & 0.5              & \textgreater0.5  & \textless $-$0.2   & ---               & ---                & 0.2           \\
&15509.737                               & ---                & ---                & 0.2              & \textgreater0.5  & ---                & \textless$-$0.2   & 0.1              & 0.3           \\
&15564.252                               & ---                & 0.2              & 0.1              & 0.3              & $-$0.4             & ---               & 0.2              & \textless$-$0.2 \\

	\end{tabular}
        	\end{threeparttable}
\end{table*}

\begin{table*}
	\centering
	\contcaption{}
    	\begin{threeparttable}
	\begin{tabular}{llllllllll} 
		\hline
	Element & Lambda (\AA)  & Star I   & Star II  & Star D  & Star E& Star F & Star F* & Star G & Star H \\ \hline

&15573.254                               & ---                & ---                & 0.2              & 0.7              & ---                & ---               & ---                & 0.4           \\
&15576.255                               & ---                & ---                & ---                & 0.7              & \textless 0.0    & ---               & ---                & ---             \\
&15631.270                                & ---                & ---                & ---                & 0.7              & ---                & ---               & ---                & ---             \\   
&16056.386                               & 0.5              & ---                & ---                & 0.7              & ---                & ---               & ---                & $-$0.2          \\
&16065.388                               & 0.5              & 0.2              & 0.2              & $-$0.2             & $-$0.3             & 0.0               & 0.2              & $-$0.2          \\
&16069.389                               & ---                & ---                & ---                & ---                & ---                & ---               & ---                & 0.4\textless  \\
&16196.424                               & ---                & ---                & ---                & ---                & ---                & ---               & ---                & ---             \\
&16256.440                                & ---                & ---                & ---                & ---                & ---                & ---               & 0.0                & ---             \\
&16264.442                               & ---                & ---                & ---                & ---                & ---                & ---               & ---                & ---             \\
&16350.466                               & 0.5              & 0.2              & 0.5              & 0.3              & 0.2              & ---               & ---                & 0.2           \\
&16356.467                               & ---                & ---                & ---                & ---                & ---                & ---               & ---                & ---             \\
&16358.467                               & ---                & 0.5              & 0.4              & 0.3              & \textless 0.2    & ---               & ---                & ---             \\
&16372.472                               & \textgreater0.6  & ---                & 0.2              & 0.7              & ---                & \textless$-$0.2   & ---                & ---             \\
&16530.515                               & ---                & ---                & ---                & ---                & ---                & \textless$-$0.2   & ---                & ---             \\
&16539.517                               & \textgreater0.6  & ---                & \textgreater0.2  & 0.7              & ---                & ---               & ---                & 0.2           \\
&16543.518                               & ---                & ---                & ---                & ---                & ---                & 0.0               & 0.2              & ---             \\
&16708.563                               & 0.5              & ---                & 0.2              & 0.7              & ---                & \textless$-$0.2   & ---                & 0.4           \\
&16718.566                               & \textgreater0.5  & 0.5              & 0.5              & 0.7              & ---                & \textless$-$0.2   & ---                & 0.4           \\
&                  &                  &                  &                  &                  &                 &                  &               \\

C from CO lines       &   15325.187  & ---              & $<$0.1                & 0.1             & ---              & ---             & ---            & ---                  & 0.2           \\
    &   15367.243                  & ---              			& ---                        & $<$ 0.2	& 0.1              & 0.2             & ---            & $<$ 0.2              & $-$0.1           \\
    &   15470.226                  & ---              		  & ---                          & ---             & ---                 & 0.4             & ---            & ---              & 0.2          \\
    &   15499.690                  & ---              		   & $<$0.0                & ---             & ---                 & ---             & ---            & ---                 & 0.2           \\
    &   16186.421                  & $<$ 0.0                & $<$0.2                & 0.1             & 0.0            & ---             & ---            & ---                 & ---           \\
    &   16193.923                  & ---        			        & ---                        & ---             & ---                  & 0.2             & ---            & ---              & 0.0           \\

\\
MgI       &   15745.017                  & 0.0                & 0.0                & $-$0.2             & 0.1              & $-$0.2             & $-$0.2            & 0.8              & 0.5           \\
&15753.189                               & $-$0.2             & $-$0.1             & $-$0.1             & 0.3              & $-$0.5             & $-$0.6            & 0.8              & 0.5           \\
&15770.150                                & $-$0.3             & $-$0.2             & $-$0.4             & 0.3              & ---                & $-$0.6            & 0.7              & 0.5           \\
&15958.836                               & ---                & 0.1              & 0.3              & 0.2              & $-$0.2             & ---               & \textless 0.4    & 0.3           \\
&                  &                  &                  &                  &                  &                 &                  &               \\
&                  &                  &                  &                  &                  &                 &                  &               \\
MnI  &       15222.009                   & $-$0.1             & \textless 0.0    & $-$0.2             & 0.1              & 0.4              & 0.3             & 0.2              & 0.2           \\
&15791.657                               & ---                & ---                & 0.0                & 0.0                & ---                & \textless0.0     & ---                & ---             \\
&15969.543                               & ---                & $-$0.1             & $-$0.2             & $-$0.2             & 0.1              & 0.2             & $-$0.3             & $-$0.1          \\
&                  &                  &                  &                  &                  &                 &                  &               \\
CaI   &     16161.778                    & 0.2              & 0.4              & 0.3              & 0.2              & $-$0.2             & ---               & ---                & 0.1           \\
&16208.514                               & \textless0.0     & 0.0                & 0.4              & 0.3              & $-$0.4             & ---               & ---                & ---             \\
&16213.018                               & ---                & ---                & 0.0                & 0.0                & ---                & ---               & ---                & ---             \\
&                  &                  &                  &                  &                  &                 &                  &               \\
SI   &        15426.404                  & 0.2              & \textless0.1     & 0.0                & 0.6              & ---                & ---               & ---                & 0.2           \\
&15474.047                               & ---                & ---                & ---                & 0.0                & $-$0.3             & $-$0.2            & ---                & 0.1           \\
&15482.710                                & 0.1              & 0.2              & 0.2              & $-$0.2             & $-$0.5             & ---               & ---                & 0.3           \\
&                  &                  &                  &                  &                  &                 &                  &               \\
&                  &                  &                  &                  &                  &                 &                  &               \\
AlI  & 16723.527                         & $-$0.1             & $-$0.1             & 0.4              & 0.3              & 0.8              & 0.8             & \textless0.7     & 0.3           \\
&16767.939                               & 0.2              & \textless0.2     & 0.4              & 0.3              & 0.7              & 0.8             & ---                & 0.3           \\
&                  &                  &                  &                  &                  &                 &                  &               \\
KI  &   15172.521                        & 0.1              & 0.2              & $-$0.2             & 0.0                & 0.1              & $-$0.2            & 0.7              & \textgreater0.0 \\
&                  &                  &                  &                  &                  &                 &                  &               \\
NaI   &  16378.326                       & ---                & ---                & ---                & \textgreater0.1  & ---                & ---               & ---                & ---             \\
&16393.327                               & ---                & ---                & ---                & ---                & \textgreater$-$0.4 & \textless$-$0.4   & ---                & ---             \\
&                  &                  &                  &                  &                  &                 &                  &               \\
SiI    &      15562.031                  & \textless$-$0.2    & 0.2              & ---                & 0.2              & ---                & ---               & 0.2              & ---             \\
&16055.585                               & ---                & ---                & ---                & ---                & ---                & ---               & ---                & ---             \\
&16064.397                               & $-$0.2             & 0.0                & 0.0                & $-$0.2             & ---                & ---               & ---                & 0.0             \\
&16099.184                               & \textless$-$0.2    & 0.2              & 0.4              & 0.0                & $-$0.4             & $-$0.4            & \textless0       & 0.2           \\
&16185.583                               & $-$0.2             & ---                & 0.0                & 0.2              & ---                & ---               & 0.0                & 0.1           \\
&16191.217                               & ---                & ---                & ---                & ---                & ---                & ---               & ---                & ---             \\
&16220.100                                 & 0.0                & 0.2              & $-$0.1             & $-$0.2             & $-$0.4             & $-$0.4            & 0.0                & 0.2           \\
&16685.327                               & 0.2              & 0.2              & 0.2              & 0.0                & 0.4              & 0.4             & 0.2\textless     & 0.2           \\
&16832.756                               & 0.0                & ---                & 0.2              & 0.0                & 0.4              & 0.4             & 0.2              & 0.0     \\

		\hline
	\end{tabular}
                \begin{tablenotes}
        \item[\textit{a}] [X/H] is given instead for Fe I.
        \end{tablenotes}
	\end{threeparttable}
\
\end{table*}






\bsp	
\label{lastpage}
\end{document}